\documentclass[prd,
superscriptaddress,preprintnumbers,tightenlines,nofootinbib, eqsecnum]{revtex4-2}

\usepackage{amsmath}
\usepackage{amsfonts}
\usepackage{amssymb}
\usepackage{bm}
\usepackage{hyperref}
\usepackage{mathrsfs}
\usepackage{graphicx}
\usepackage{empheq}
\usepackage{ulem}
\usepackage{bbold}
\usepackage[usenames]{color}
\usepackage{ytableau}

\usepackage{subcaption}
\usepackage{booktabs}

\allowdisplaybreaks

\DeclareSymbolFontAlphabet{\mathrsfs}{rsfs}
\DeclareMathAlphabet{\mathcal}{OMS}{cmsy}{m}{n}

\newcommand{\nn}{\nonumber}
\newcommand\calO{\mathcal{O}}
\newcommand{\dd}{\mathrm{d}}
\newcommand{\di}{\mathrm{i}}

\newcommand{\tr}{\mathrm{Tr}}

\newcommand{\calD}{\mathcal{D}}

\newcommand{\dtc}[1]{{\Bigl(\frac{\dd{#1}}{\dd t}\Bigr)}^{2}}

\newcommand{\rhoa}[1]{\mathop{\rho}_{#1}}
\newcommand{\rhotaua}[1]{\mathop{\rho}_{#1}{}_{\!\tau}}
\newcommand{\uau}[2]{\mathop{u}_{#1}{}^{\!{#2}}}
\newcommand{\ual}[2]{\mathop{u}_{#1}{}_{\!{#2}}}
\newcommand{\pau}[2]{\mathop{p}_{#1}{}^{\!{#2}}}
\newcommand{\pal}[2]{\mathop{p}_{#1}{}_{\!{#2}}}

\newcommand{\pdotal}[2]{\mathop{\dot{p}}_{#1}{}_{\!{#2}}}

\newcommand{\Piau}[2]{\mathop{\Pi}_{#1}{}^{\!{#2}}}
\newcommand{\Pihatau}[2]{\mathop{\hat{\Pi}}_{#1}{}^{\!{#2}}}
\newcommand{\Pial}[2]{\mathop{\Pi}_{#1}{}_{\!{#2}}}

\newcommand{\Jiau}[2]{\mathop{\mathcal{J}}_{#1}{}^{\!{#2}}}
\newcommand{\Jihatau}[2]{\mathop{\hat{\mathcal{J}}}_{#1}{}^{\!{#2}}}
\newcommand{\Hau}[2]{\mathop{H}_{#1}{}^{\!{#2}}}
\newcommand{\Haul}[3]{\mathop{H}_{#1}{}^{\!{#2}}_{~{#3}}}
\newcommand{\Hal}[2]{\mathop{H}_{#1}{}_{\!{#2}}}
\newcommand{\Fau}[2]{\mathop{F}_{#1}{}^{\!{#2}}}
\newcommand{\Fal}[2]{\mathop{F}_{#1}{}_{\!{#2}}}
\newcommand{\Qau}[2]{\mathop{Q}_{#1}{}^{\!{#2}}}

\newcommand{\Kal}[2]{\mathop{K}_{#1}{}_{\!{#2}}}
\newcommand{\Hduala}[2]{\mathop{\stackrel{\star}{H}}_{#1}{}^{\!{#2}}}
\newcommand{\perpa}[3]{\mathop{\perp}_{#1}{}^{\!{#2}}_{\!{#3}}}

\newcommand{\xiau}[2]{\mathop{\xi}_{#1}{}^{\!{#2}}}
\newcommand{\xiperpau}[2]{\mathop{\xi}_{#1}{}^{\!{#2}}_{\!\perp}}
\newcommand{\xiperpal}[2]{\mathop{\xi}_{#1}{}_{\!\perp{#2}}}
\newcommand{\xidotperpau}[2]{\mathop{\dot{\xi}}_{#1}{}^{\!{#2}}_{\!\perp}}
\newcommand{\xidotperpal}[2]{\mathop{\dot{\xi}}_{#1}{}_{\!\perp{#2}}}
\newcommand{\xiddotperpau}[2]{\mathop{\ddot{\xi}}_{#1}{}^{\!{#2}}_{\!\perp}}
\newcommand{\xiddotperpal}[2]{\mathop{\ddot{\xi}}_{#1}{}_{\!\perp{#2}}}

\newcommand{\Eau}[2]{\mathop{E}_{#1}{}^{\!{#2}}}
\newcommand{\Eal}[2]{\mathop{E}_{#1}{}_{\!{#2}}}

\newcommand{\phia}[1]{\mathop{\phi}_{#1}}
\newcommand{\etaa}[1]{\mathop{\eta}_{#1}}
\newcommand{\ma}[1]{\mathop{m}_{#1}}

\newcommand{\Aal}[2]{\mathop{A}_{#1}{}_{\!{#2}}}

\newcommand{\Dal}[2]{\mathop{D}_{#1}{}_{\!{#2}}}

\newcommand{\vau}[2]{\mathop{v}_{#1}{}^{\!{#2}}}

\newcommand{\ta}[1]{\mathop{t}_{#1}}

\newcommand{\bmDal}[1]{\mathop{\bm{D}}_{#1}}

\newcommand{\bmXa}[1]{\mathop{\bm{X}}_{#1}}
\newcommand{\bmYa}[1]{\mathop{\bm{Y}}_{#1}}

\newcommand{\bmxa}[1]{\mathop{\bm{x}}_{#1}}
\newcommand{\bmya}[1]{\mathop{\bm{y}}_{#1}}

\newcommand{\bmEa}[1]{\mathop{\bm{E}}_{#1}}

\newcommand{\bmAa}[1]{\mathop{\bm{A}}_{#1}}

\newcommand{\bmxiperpa}[1]{\mathop{\bm{\xi}}_{#1}{}_{\!\perp}}

\newcommand{\Jau}[2]{\mathop{J}_{#1}{}^{\!{#2}}}
\newcommand{\Jetoileau}[2]{\mathop{J}_{#1}{}_*^{\!{#2}}}
\newcommand{\jau}[2]{\mathop{j}_{#1}{}^{\!{#2}}}
\newcommand{\jbarau}[2]{\mathop{\overline{j}}_{#1}{}^{\!{#2}}}
\newcommand{\jetoileau}[2]{\mathop{j}_{#1}{}_*^{\!{#2}}}
\newcommand{\jetoilebarau}[2]{\mathop{\overline{j}}_{#1}{}_*^{\!{#2}}}

\newcommand{\Jstarau}[2]{\mathop{J}_{#1}{}_{\!*}^{\!{#2}}}
\newcommand{\rhostara}[1]{\mathop{\rho}_{#1}{}_{\!\!*}}
\newcommand{\omegaa}[1]{\mathop{\omega}_{#1}}

\newcommand{\Qcal}{{\cal Q}}
\newcommand{\Tcal}{{\cal T}}

\newcommand{\be}{\begin{equation}}
\newcommand{\ee}{\end{equation}}
\newcommand{\bse}{\begin{subequations}}
\newcommand{\ese}{\end{subequations}}

\definecolor{darkgreen}{rgb}{0,0.5,0}

\hypersetup{
 unicode=false,          
 pdftoolbar=true,        
 pdfmenubar=true,        
 pdffitwindow=false,     
 pdfstartview={FitH},    
 pdftitle={My title},    
 pdfauthor={Author},     
 pdfsubject={Subject},   
 pdfcreator={Creator},   
 pdfproducer={Producer}, 
 pdfkeywords={keyword1} {key2} {key3}, 
 pdfnewwindow=true,      
 colorlinks=true,       
 linkcolor=red,          
 citecolor=cyan,        
 filecolor=magenta,      
 urlcolor=darkgreen,           
 linktocpage=true
}

\makeatletter
\g@addto@macro\bfseries{\boldmath}
\makeatother

\begin{document}
	
\title{Effective field theory reproducing the MOND phenomenology \\based on a non-Abelian Yang-Mills graviphoton}

\author{Luc \textsc{Blanchet}}\email{luc.blanchet@iap.fr}
\affiliation{$\mathcal{G}\mathbb{R}\varepsilon{\mathbb{C}}\calO$, 
	Institut d'Astrophysique de Paris, UMR 7095, CNRS, \\Sorbonne Universit{\'e},
	98\textsuperscript{bis} boulevard Arago, 75014 Paris, France}

\author{Emeric \textsc{Seraille}}\email{emeric.seraille@phys.ens.fr}
\affiliation{Laboratoire de Physique de l’Ecole Normale Sup{\'e}rieure, ENS, CNRS, Universit{\'e} \\ PSL, Sorbonne Universit{\'e}, Universit{\'e} Paris Cit{\'e}, F-75005 Paris, France
}
\affiliation{$\mathcal{G}\mathbb{R}\varepsilon{\mathbb{C}}\calO$, 
Institut d'Astrophysique de Paris, UMR 7095, CNRS, \\Sorbonne Universit{\'e},
98\textsuperscript{bis} boulevard Arago, 75014 Paris, France}

\date{\today}

\begin{abstract}
Motivated by the phenomenology of MOND, we propose a theory based on a fundamental non-Abelian Yang-Mills gauge field with gravitational coupling constant (a ``graviphoton'') emerging in a regime of weak acceleration, i.e. below the MOND acceleration scale. Using the formalism of the effective field theory and invoking a mechanism of gravitational polarization of the dark matter medium, we show that generic solutions of this theory reproduce the deep MOND limit without having to introduce in an ad hoc way an arbitrary function in the action. In this framework, MOND is due to the existence of a new sector of the standard model of particle physics. Furthermore, the model involves a violation of the local Lorentz invariance in the low acceleration regime. We show how to restore the general covariance of the model by adding one gravitational degree of freedom in the form of the scalar Khronon field. 
\end{abstract}

\pacs{04.25.Nx, 04.30.-w, 97.60.Jd, 97.60.Lf}

\maketitle


\section{Physical motivations}
\label{sec:intro}

The phenomenology of the Modified Newtonian Dynamics (MOND)~\cite{Milg1, Milg2, Milg3} could be a key to understand the mystery of dark matter. This phenomenology is encoded in the MOND empirical formula, which is able to reproduce in great detail many aspects of the distribution and dynamics of dark matter at the scale of galaxies, including the flat rotation curves of galaxies, the baryonic Tully-Fisher relation for spirals, and the tight correlation between the presence of dark matter and the level of acceleration below the critical acceleration scale $a_0\simeq 1.2\times 10^{-10}\,\text{m}\,\text{s}^{-2}$ (Milgrom's law). Such observations taking place at the scale of galaxies remain essentially unexplained by the standard cosmological model $\Lambda$-CDM based on cold dark matter and the cosmological constant. In particular, the standard model falls short in explaining in a natural way the Milgrom law (see~\cite{FamMcG12} for a comprehensive review). On the other hand, the MOND formula faces important challenges when extrapolated either at the smaller scale of the solar system~\cite{Milg09, BN11, HFAG15, VNT24} or at the larger scale of clusters of galaxies~\cite{GD92, Sand99, PS05}.   

Relativistic extensions of general relativity (GR) recovering the MOND phenomenology in the weak field, weak acceleration regime have been proposed, which purport to fit observations using additional gravitational fields but without dark matter. The classic example is the Tensor-Vector-Scalar (TeVeS) theory~\cite{Sand97, Bek04, Sand05}, which is, however, ruled out by cosmological observations~\cite{SMFB06}. Many other MOND-motivated GR extensions without dark matter have been proposed~\cite{BGef07, ZFS07, HZL08, LBMZ08, Sk08, bimond1, Zu10, BDgef11, DEW11, BM11, Sand11, Mendoza12, Woodard14, Milgrom19, DAmbrosio20, SZ21, BS24}. The state-of-the-art today is theories able to reproduce the MOND limit in galaxies and also to recover the dark matter we observe in cosmological scales, at first order in cosmological perturbation~\cite{SZ21, BS24, HN24}. We note that most of the previous theories cannot be considered as ``fundamental'', as they need an arbitrary function in the action which is introduced by hands in order to fit the MOND interpolating function, and which is not explained by fundamental physics. 

A different class of MOND theories, called modified dark matter or sometimes hybrid models, attempts at modifying the properties of dark matter rather than changing the gravitational law (namely GR). In this class are models of dark matter superfluidity~\cite{BK15, BK16, BK25review} and dipolar dark matter. Initially formulated in the Newtonian approximation~\cite{B07mond}, the dipolar dark matter (DDM) model was then extended to a relativistic theory~\cite{BL08, BL09}, which was proved to be in agreement with the cosmological model $\Lambda$-CDM at first order in cosmological perturbation. However, the relativistic theory~\cite{BL08, BL09} has some drawbacks, notably the presence of an instability in the dark matter sector, and still contains an ad hoc arbitrary function matched to the MOND interpolating function without physical justification.

The DDM model is physically motivated by the striking \textit{dielectric} analogy of MOND~\cite{B07mond}. The non-relativistic field equation for MOND takes the form of a modification of the Poisson equation for the gravitational field given by~\cite{BekM84}
\begin{equation}
	\label{Poisson_milgrom}	\bm{\nabla}\cdot\left[\mu\Bigl(\frac{g}{a_0}\Bigr) \bm{g} \right] = - 4 \pi G \rho_\text{bar}\,, 
\end{equation}
where $\bm{g} = \bm{\nabla} U$ with $U$ the ordinary Newtonian potential, $\rho_\text{bar}$ is the mass density of baryons (stars and gas in a galaxy) and $\mu(g/a_0)$ is the MOND interpolating function of the ratio between $g=\vert\bm{g}\vert$ and the MOND acceleration scale $a_0$. In the strong acceleration limit $g\gg a_0$ we have $\mu\simeq 1$ and we recover the usual Poisson equation. In the weak acceleration regime $g \ll a_0$ we have $\mu\simeq g/a_0$ which describes the so-called deep MOND regime. It is remarkable that this simple behavior of the interpolating function when $g \ll a_0$ permits to predict so many aspects of the dark matter distribution in galaxies~\cite{Milg1, Milg2, Milg3}. Various families of interpolating functions $\mu(g/a_0)$ have been proposed (but without clear derivation from first principles) and tested against astronomical observations~\cite{FamMcG12}.

The MOND equation~\eqref{Poisson_milgrom} is analogous to the Gauss law describing the electric field inside a polarizable dielectric medium, namely $\bm{\nabla}\cdot\bm{D}=\rho_\text{free}$ where $\rho_\text{free}$ is the density of free charges, and the electric displacement field $\bm{D}$ (or induction) differs from the applied external field $\bm{E}$ due to the polarization of the medium, $\bm{D}=\epsilon_0\bm{E}+\bm{\Pi}_e$ where $\bm{\Pi}_e$ is the polarization related to the charges bounded to the atoms in the dielectric as $\rho_\text{bound} = - \bm{\nabla}\cdot\bm{\Pi}_e$. As a first approximation, the polarization is proportional to the electric field and we can introduce the electric susceptibility coefficient $\chi_e$ such that $\bm{\Pi}_e=\epsilon_0\chi_e\bm{E}$. In more general nonlinear dielectric media, the susceptibility is a function of the norm of the electric field, $\chi_e(E)$. 

In keeping with the analogy, we can write the MOND interpolating function as $\mu=1+\chi$ where $\chi=\chi(g)$ is interpreted as a coefficient of gravitational susceptibility. The gravitational polarization reads then
\begin{equation}
	\label{Polarization}
	\bm{\Pi} = - \frac{\chi}{4\pi G} \,\bm{g}\,, 
\end{equation}
while the analogue of the bounded charges in electrostatics can naturally be interpreted as dark matter defined with 
\begin{equation}
	\label{rhoDM}
	\rho_\text{DM} = - \bm{\nabla}\cdot\bm{\Pi}\,. 
\end{equation}
In Ref.~\cite{B07mond} a microscopic description of the DDM medium based on gravitational dipoles was proposed, which confirmed the relevance of this interpretation for MOND. In particular, it was shown that while in the electric case the electric susceptibility is positive ($\chi_e>0$), corresponding to the screening of electric fields by polarization charges, in the gravitational case we should have $\chi<0$. Consequently, there is an \textit{anti-screening} of ordinary masses by polarization masses, hence enhancement of the gravitational field, which leads to a dark matter effect consistent with MOND (since $0<\mu<1$). It was furthermore shown that the DDM medium requires an internal non-gravitational force (a ``fifth'' force) linking together the microscopic constituents of the dipole in order for the MOND equation to be recovered at equilibrium between the internal force and the gravitational force. An obvious creepy aspect of the model is that the gravitational dipole should be made of particles with positive and negative gravitational masses $m_\text{g}$ (the inertial masses $m_\text{i}$ are always positive). If we accept this, the DDM medium appears like a globally neutral plasma made of positive and negative masses, with stable plasma-like oscillations around equilibrium~\cite{B07mond}. 

In this paper, we present a novel approach to the DDM model. We propose that the internal non-gravitational fifth force of this theory is a non-Abelian Yang-Mills (YM) gauge field based on the gauge group $\text{SU}(2)$. Since the internal field must be able to cancel the gravitational force, the coupling constant $g$ of the usual YM quadratic term is assumed to be linked to the gravitational constant $G$, \textit{i.e.} scaling with $G$ as $g\propto\sqrt{G}$. Furthermore we assume that this is valid at the specific energy scale corresponding to the deep-MOND regime. We shall coin this gravitational-related internal vector field a (non-Abelian version of) ``graviphoton''~\cite{Scherk}.\footnote{The name ``graviphoton'' was proposed by P. Fayet.} We shall show how this field can be coupled to the DDM medium, described by a specific polarization tensor. The motivation behind the choice of non-Abelian YM field is that we shall be able, following arguments from the effective field theory (EFT), to describe the deep MOND regime without invoking an arbitrary function in the action. Furthermore the cosmological constant comes naturally along from the EFT. We shall regard the fifth force as a new interaction emerging in the weak-field small-acceleration regime $\vert\bm{g}\vert\ll a_0$, where $\bm{g}$ is the Newtonian gravitational field, and representing a new sector of particle physics, supposed to be unified with the other interactions of the standard model described by the product of gauge groups $\text{U}(1)\times\text{SU}(2)\times\text{SU}(3)$ (a related approach, with similar motivation, is described in~\cite{Finster_2024}). However, in the present state of the model, we cannot describe the transition between the deep MOND regime and the ordinary Newtonian regime. This transition may require a non-perturbative formulation in the strong field regime using the theory of effective Lagrangians~\cite{EulerHeisenberg, Dunne05, ItzyksonZuber}. 

The DDM medium is described by gravitational dipole moments made of doublets of positive and negative gravitational masses\footnote{The precise nature of the particles forming the dipoles (fermions or bosons) does not need to be specified at this stage; the physical consequences discussed in this paper are only associated to the existence of a dipolar medium subject to gravitational polarization.} interacting both gravitationally and through the Yang-Mills internal field. More precisely, we consider three doublets of particles with
\begin{align}\label{doublet}
	\bigl(\ma{a}{}_\text{i}, \ma{a}{}_\text{g}\bigr)=\bigl(\ma{a},\pm \ma{a}\bigr)\,,
\end{align}
where $a=1,2,3$ is the internal $\text{SU}(2)$ index. Because of the negative gravitational masses, there is a violation of the weak equivalence principle. As a consequence, the dipolar particles are accelerated not by the gravitational field but by its gradient namely the tidal gravitational field. The doublets of particles~\eqref{doublet} can be used to build three gravitational dipoles. Moreover, we shall make a distinction between their gravitational masses when they are coupled to gravity or to the YM field [see Eq.~\eqref{defeta}].

In a Newtonian model with gravitational dipoles, one can write an action which will contain a coupling between the dipole moments $\bm{\xi}$ (or polarization vector $\bm{\Pi}=\rho\,\bm{\xi}$) and the Newtonian gravitational field $\bm{g}$. Such a coupling $\bm{\Pi}\cdot\bm{g}$ in the theory's action, which is the analogue of the usual $\bm{\Pi}_e\cdot\bm{E}$ in electromagnetism, cannot be written in a covariant way, since the gravitational field $\bm{g}$ can always be gauged away by going to a freely falling frame. As a result, the general covariance (or diffeomorphism invariance) of the theory is lost and in particular there will be a violation of the local Lorentz invariance (LLI). This violation occurs only in the regime of very weak accelerations, below the MOND scale $a_0$. The breakdown of diffeomorphism invariance at low accelerations is similar to the one advocated in relativistic MOND-motivated Khronon theories~\cite{BM11, Sand11, BS24} (which were partly inspired by the Ho{\v{r}}ava gravity, itself motivated by quantum gravity~\cite{Horava2009, Blas2009, Blas2011}) and has also been seen by Milgrom~\cite{Milgrom19} as a route for MOND. 

Finally, we shall show that the DDM model can be promoted to a covariant theory by adding the Khronon scalar field, which labels a foliation of space-time by space-like three-dimensional hypersurfaces. With the Khronon field we can define the covariant space-like acceleration associated with the foliation. Since this acceleration reduces to the ordinary gravitational field $\bm{g}$ in the Newtonian limit, this gives the possibility of a covariant description of the coupling between the gravitational field and the dipole moment, therefore mimicking the presence of negative masses as in~\eqref{doublet}. The argument uses the unitary gauge for the Khronon, which is compatible with the standard post-Newtonian ansatz for the metric only for stationary systems. Thus, we find the same limitation to stationary systems as in recent Khronon theories (see discussions in~\cite{Flanagan23,BS24}). The fact that we recover MOND only for stationary systems is still satisfying, because most MOND tests are performed in stationary situations. Alternatively, one may accept the violation of the LLI and build a $3$+$1$ model in which MOND will be valid even for non-stationary systems. 

The plan of this paper is organized as follows. In Sec.~\ref{sec:YM}, we recapitulate the Yang-Mills gauge field formalism (in an arbitrary background) in the context of the effective theory, and adapt it to the present case of the coupling to an antisymmetric polarization tensor describing the DDM medium. In Sec.~\ref{sec:Newtonian} we describe a complete model at the Newtonian level for the DDM particles and their interactions with gravity and the YM field, and show that the generic solution of the equations naturally yields the MOND limit. In Sec.~\ref{sec:covariant} we generalize the Newtonian model to a relativistic theory based on the Khronon scalar field, we obtain the stress-energy tensor of the dark matter and we check the consistency with the relativistic equations. Finally, we summarize the paper and discuss some future directions in Sec.~\ref{sec:conclusion}. A toy model for dipolar particles is relegated to Appendix~\ref{microscopic}.

\section{The Yang-Mills gauge field and effective theory}
\label{sec:YM}

\subsection{General formalism}

The central aspect of the DDM model is the internal interaction between the constituents of the dipole moments, chosen here to be a non-Abelian Yang-Mills (YM) gauge field associated with the group $\text{SU}(2)$ of local invariance.\footnote{We follow the conventions of Ref.~\cite{ItzyksonZuber}. The generators of the Lie algebra are denoted $t_a$ with $a, b, c, \dots = 1, 2, 3$. The generators are anti-hermitian ($t_a^\dag=-t_a$) and given by $t_a=\frac{\di}{2}\sigma_a$ where $\sigma_a$ are the usual $2\times 2$ Pauli matrices. Thus, the generators are normalized according to $\tr(t_a t_b)=-\frac{1}{2}\delta_{ab}$ and satisfy the commutation relations $[t_a, t_b] = -\sum_c \epsilon_{abc} \,t_c$ where $\epsilon_{abc}$ is the totally antisymmetric Levi-Civita symbol with $\epsilon_{123}=1$. For clarity we explicitly indicate the summation symbols over the internal $\text{SU}(2)$ indices (while we keep Einstein's summation convention for the repeated ordinary spatial indices $i$, $j$, $\cdots$).} The gauge covariant derivative is
\begin{align}\label{covder}
	\mathcal{D}_\mu = \mathbb{1} \nabla_\mu  + \frac{1}{\ell} K_\mu\,,
\end{align}
where $\nabla_\mu$ denotes the covariant derivative associated with the metric $g_{\mu\nu}$ with signature $(-,+,+,+)$, $\ell$ is a constant having the dimension of a length, and the gauge field $K_\mu$ is an element of the Lie algebra, 
\begin{align}
	K_\mu = \sum_a \mathop{K}_{a}{}_{\!\!\mu} \,\ta{a}\,,
\end{align}
chosen to be dimensionless. The field strength tensor (having the dimension of an inverse length) is defined by
\begin{subequations}
	\begin{align}
		H_{\mu\nu} \equiv \mathcal{D}_\mu K_\nu - \mathcal{D}_\nu K_\mu = \partial_\mu K_\nu - \partial_\nu K_\mu + \frac{1}{\ell}\bigl[K_\mu, K_\nu\bigr] = \ell\,\bigl[\mathcal{D}_\mu, \mathcal{D}_\nu\bigr]
		\,,
	\end{align}
	or, equivalently, in components,
	\begin{align}
		\mathop{H}_{a}{}_{\!\!\mu\nu} = \partial_\mu \!\mathop{K}_{a}{}_{\!\!\nu} - \partial_\nu \!\mathop{K}_{a}{}_{\!\!\mu} + \frac{1}{\ell} \,\sum_{b,c}\epsilon_{abc} \mathop{K}_{b}{}_{\!\!\mu}\mathop{K}_{c}{}_{\!\!\nu}\,.
	\end{align}
\end{subequations}

In this paper, we shall couple the YM field to a classical current $\hat{\mathcal{J}}^\mu$, belonging to the Lie algebra, which is conserved with respect to the gauge covariant derivative,
\begin{align}\label{consJmu}
	\bigl[ \calD_\mu, \hat{\mathcal{J}}^{\mu}\bigr] = \nabla_\mu \hat{\mathcal{J}}^\mu + \frac{1}{\ell} \bigl[ K_\mu, \hat{\mathcal{J}}^\mu\bigr] = 0\, .
\end{align}
Furthermore, we shall consider the case where this current is the covariant derivative of an antisymmetric tensor field $\hat{\Pi}^{\mu\nu}$ called the polarization tensor (also a member of the Lie algebra), in the sense that
\begin{align}\label{Jmu}
	\hat{\mathcal{J}}^\mu \equiv - c \bigl[ \mathcal{D}_\nu, \hat{\Pi}^{\mu\nu}\bigr] = - c \Bigl(\nabla_\nu \hat{\Pi}^{\mu\nu} + \frac{1}{\ell} \bigl[ K_\nu, \hat{\Pi}^{\mu\nu}\bigr]\Bigr) \,.
\end{align}
We introduce a factor $c$ so that $\hat{\mathcal{J}}^\mu$ and $\hat{\Pi}^{\mu\nu}$ have the usual dimensions of a current and a polarization; our hat notation will be clarified later in Eq.~\eqref{Pihat}. The polarization tensor $\hat{\Pi}^{\mu\nu}$ represents the independent degrees of freedom of the matter field, i.e. the DDM medium. It is thus independent of the YM gauge field $K_\mu$ and as such is expected to commute with the field strength, namely
\begin{align}\label{commPiH}
	\bigl[\hat{\Pi}^{\mu\nu}, H_{\mu\nu}\bigr] = 0\,.
\end{align}
It is straightforward to check that under this assumption the current~\eqref{Jmu} indeed satisfies the conservation law~\eqref{consJmu}.

Next, we define the gauge invariant action of the YM field coupled to the polarization tensor $\hat{\Pi}^{\mu\nu}$. Following the EFT, we must include in the action all possible terms compatible with the symmetries and the gauge invariance. In the Abelian case the action is quadratic in the field strength and higher order corrections appear at quartic level~\cite{EulerHeisenberg}. In the non-Abelian case, however, there is a possibility, besides the standard quadratic term, of a cubic term in the action. For the present purpose we adopt the term $\sim \tr(H\,H\,H^\star)$ (considered in e.g.~\cite{PhysRevLett.63.2333} for the $\text{SU}(3)$ gauge group) built from the dual tensor\footnote{The Levi-Civita tensor is defined by $\varepsilon^{\mu\nu\rho\sigma} = - \frac{1}{\sqrt{-g}} \epsilon^{\mu\nu\rho\sigma}$ and $\varepsilon_{\mu\nu\rho\sigma} = \sqrt{-g} \,\epsilon_{\mu\nu\rho\sigma}$ in terms of the completely anti-symmetric Levi-Civita symbol $\epsilon_{\mu\nu\rho\sigma}=\epsilon^{\mu\nu\rho\sigma}$ such that $\epsilon_{0123}=1$.}
\begin{align}\label{Hdual}
	\stackrel{\star}{H}{}^{\!\!\mu\nu} \equiv \varepsilon^{\mu\nu\rho\sigma}{H}_{\rho\sigma}\,.
\end{align}
In principle, in an EFT approach, we should also include the cubic term $\sim \tr(H\,H\,H)$. This term is not considered as it will not contribute to the Newtonian limit investigated in Sec.~\ref{sec:Newtonian}.

In front of the cubic term we introduce a constant $\alpha$ with the dimension of a length, which will be related later to the MOND acceleration scale $a_0$. Furthermore, we assume that the coupling constant of the YM field is directly given by the Newton constant $G$. The YM part of the action is therefore given by the Lagrangian\footnote{In our convention $K_\mu$ is dimensionless and the coupling constant $\ell$ has the dimension of a length. A more usual convention for YM fields is
	\begin{align*}
		\mathop{F}_{a}{}_{\!\!\mu\nu} = \partial_\mu \!\mathop{A}_{a}{}_{\!\!\nu} - \partial_\nu \!\mathop{A}_{a}{}_{\!\!\mu} + g \,\sum_{b,c}\epsilon_{abc} \mathop{A}_{b}{}_{\!\!\mu}\!\mathop{A}_{c}{}_{\!\!\nu}\,,\qquad L_\text{YM} = - \frac{1}{4} \hbar c \Fal{a}{\mu\nu} \!\Fau{a}{\mu\nu} + \cdots\,,
	\end{align*}
	where $A_\mu$ is the inverse of a length and the coupling constant $g$ is dimensionless. The two conventions are related by $K_\mu=g\ell A_\mu$, where $g\ell=\sqrt{4\pi}\,\ell_\text{P}$ is given by the Planck length $\ell_\text{P}=\sqrt{\frac{\hbar G}{c^3}}$.
}
\begin{align}\label{LYM}
	L_\text{YM} = \tr\biggl\{ c^2 \hat{\Pi}^{\mu\nu}
	H_{\mu\nu} + \frac{c^4}{8\pi G} \biggl[ - \frac{\Lambda}{2} + H_{\mu\nu} H^{\mu\nu} + \alpha \,H_{\mu\tau} H^{\tau}_{\phantom{\tau}\nu}\!\stackrel{\star}{H}{}^{\!\!\mu\nu} + \calO(\alpha^2) \biggr]\biggr\}\,,
\end{align}
such that the action reads $S=\int\dd^4x\sqrt{-g}L$, and the dynamical variables are $K_{\mu}$ and the independent degrees of freedom in $\hat{\Pi}^{\mu \nu}$. The matter terms and gravity will be added in Eqs.~\eqref{LagN} and~\eqref{action} below. As we discarded the term $\sim \tr(H\,H\,H)$, the remainder $\calO(\alpha^2)$ indicates all higher-order corrections (quartic, \dots) allowed by the EFT but that we do not consider in the paper. Furthermore, we have added a cosmological constant $\Lambda$, whose natural order of magnitude is related to the EFT constant $\alpha$ as
\begin{align}\label{Lambda}
	\Lambda\sim \frac{1}{\alpha^2}\,.
\end{align}
By varying the action~\eqref{LYM} with respect to the YM field $K_\mu$ we obtain the field equation
\begin{align}\label{fieldeqYM}
	\nabla_\nu Q^{\mu\nu} + \frac{1}{\ell} \bigl[ K_\nu, Q^{\mu\nu}\bigr] = \frac{4\pi G}{c^3}\hat{\mathcal{J}}^\mu\,,
\end{align}
where the current $\hat{\mathcal{J}}^\mu$ is defined by Eq.~\eqref{Jmu} and we have posed
\begin{subequations}\label{exprQ}
	\begin{align}
		Q^{\mu\nu} = H^{\mu\nu} + \frac{\alpha}{2}\Bigl( - \varepsilon^{\rho\sigma\tau[\mu}\bigl[ H^{\nu]}_{\phantom{\nu]}\rho}, H_{\sigma\tau}\bigr] + \varepsilon^{\mu\nu\rho\tau} H_{\rho\sigma} H^{\sigma}_{\phantom{\sigma}\tau}\Bigr) + \calO(\alpha^2) \,,
	\end{align}
	or, equivalently, in components,
	\begin{align}
		\Qau{a}{\mu\nu} = \Hau{a}{\mu\nu} + \frac{\alpha}{2}\sum_{b,c}\epsilon_{abc}\Bigl( \varepsilon^{\rho\sigma\tau[\mu}\Haul{b}{\nu]}{\,\rho} \Hal{c}{\sigma\tau} - \frac{1}{2}\varepsilon^{\mu\nu\rho\tau} \Hal{b}{\rho\sigma} \Haul{c}{\sigma}{\tau}\Bigr) + \calO(\alpha^2) \,.
	\end{align}
\end{subequations}
The conservation law~\eqref{consJmu} can be derived as a consequence of the field equations~\eqref{fieldeqYM}.

Thanks to the assumption that the current is the covariant derivative of the polarization, see Eq.~\eqref{Jmu}, we observe that the field equations~\eqref{fieldeqYM} can actually be exactly integrated with the simple result
\begin{align}\label{integreq}
	Q^{\mu\nu} = - \frac{4\pi G}{c^2}\,\hat{\Pi}^{\mu\nu}\,,
\end{align}
which will constitute a basic equation of the present model. It is straightforward to verify from~\eqref{exprQ} that the tensor $Q^{\mu\nu}$ commutes with $H^{\mu\nu}$, therefore the commutation relation between the polarization $\hat{\Pi}^{\mu\nu}$ and the field strength $H^{\mu\nu}$ that we have postulated in Eq.~\eqref{commPiH} is actually a consequence of the field equations.

\subsection{The polarization tensor}

The previous study has defined the coupling of the YM field to the antisymmetric polarization tensor $\hat{\Pi}^{\mu\nu}$ that describes the matter and which is a member of the $\text{SU}(2)$ Lie algebra. In turn, the polarization tensor induces the current given by~\eqref{Jmu} which also belongs to the Lie algebra and satisfies the conservation law~\eqref{consJmu}. It remains to define what are the three $\text{SU}(2)$ components of the polarization tensor in terms of some ``microscopic'' model of particles endowed with dipole moments. 

Here we implement the idea that the polarization tensor comes at a more fundamental level from three doublets of particles, with the position of the particles in each doublet being separated by the dipole moment vector. The particles in each doublet will be coupled to the YM field through the polarization and carry opposite YM charges. Hence we define (for each value of the internal index $a$) two microscopic currents, which are conserved in the usual space-time covariant sense, say
\begin{align}\label{defja}
	\jau{a}{\mu}~\text{and}~\jbarau{a}{\mu}\,,\quad\text{such that}\quad\nabla_\mu\jau{a}{\mu} = \nabla_\mu\jbarau{a}{\mu} = 0\,.
\end{align}
These currents are assumed to differ perturbatively from three conserved currents
\begin{align}\label{defJa}
	\Jau{a}{\mu} = c \rhoa{a}\uau{a}{\mu}\,,\quad\text{such that}\quad\nabla_\mu\Jau{a}{\mu} = 0\,,
\end{align}
where $\rho_a$ represents the conserved scalar mass density and $u^\mu_a$ the unit four-velocity of the dipolar particles (such that $g_{\mu\nu}\,u^\mu_a u^\mu_a=-1$). Following the procedure in Appendix B of~\cite{BB14}, we suppose that the fluid~\eqref{defJa} is made of particles with coordinate density $\rho_a^*(\mathbf{x},t)=\sum m \,\delta[\mathbf{x}-\bm{x}_a(t)]$ (with $\delta$ the usual three-dimensional Dirac function), thus satisfying the continuity equation $\partial_t \rho_a^* + \partial_i(\rho_a^* v_a^i) = 0$, where $v_a^i(\mathbf{x},t)$ is the Eulerian velocity field. Then the coordinate density of the microscopic fluid in~\eqref{defja} is defined as $r_a^*(\mathbf{x},t)=\sum m\,\delta[\mathbf{x}-\bm{x}_a(t)-\frac{1}{2}\bm{\xi}_a(t)]$, where $\xi_a^i(t)$ is the dipole moment seen here as a displacement of the position of the particles. Similarly, the coordinate density of the second microscopic fluid in~\eqref{defja} is $\overline{r}_a^*(\mathbf{x},t)=\sum m\,\delta[\mathbf{x}-\bm{x}_a(t)+\frac{1}{2}\bm{\xi}_a(t)]$. Introducing the Eulerian dipole moment field $\xi_a^i(\mathbf{x},t)$ associated with $\xi_a^i(t)$, we find that the coordinate densities and coordinate velocity fields are related (to first order in the dipole moment) by
\begin{subequations}\label{relcurrents}
	\begin{align}
		\mathop{r}_{a}{}^* &= \mathop{\rho}_{a}{}^* - \frac{1}{2}\partial_i\bigl(\mathop{\rho}_{a}{}^* \!\xiau{a}{i}\bigr) + \calO\left(\xi^2\right)\,,\\
		\mathop{w}_{a}{}^i &= \vau{a}{i}+\frac{1}{2}\frac{\dd \xi_a^i}{\dd t} - \frac{1}{2}\xiau{a}{j}\partial_j \vau{a}{i} + \calO\left(\xi^2\right)\,,
	\end{align}
\end{subequations}
and similarly for $\overline{r}_a^*$ and $\overline{w}_a^i$, where $\dd/\dd t$ is the convective derivative. Introducing the coordinate currents $j_{a*}^\mu=r_a^* w_a^\mu$, $\overline{j}_{a*}^\mu=\overline{r}_a^* \overline{w}_a^\mu$ and $J_{a*}^\mu=\rho_a^* v_a^\mu$ the previous relations become
\begin{subequations}\label{displacements*}
	\begin{align}
		\jetoileau{a}{\mu} &= \Jetoileau{a}{\mu} + \frac{1}{2}\,\partial_\nu \bigl(\Jetoileau{a}{\nu}\xiau{a}{\mu} - \Jetoileau{a}{\mu}\xiau{a}{\nu}\bigr) + \calO\left(\xi^2\right)\,,\\
		\jetoilebarau{a}{\mu} &= \Jetoileau{a}{\mu} - \frac{1}{2}\,\partial_\nu \bigl(\Jetoileau{a}{\nu}\xiau{a}{\mu} - \Jetoileau{a}{\mu}\xiau{a}{\nu}\bigr) + \calO\left(\xi^2\right)\,.
	\end{align}
\end{subequations}
Finally the expressions~\eqref{displacements*} are ``covariantized'' by defining $j_a^\mu=j_{a*}^\mu/\sqrt{-g}$, etc., and we obtain (see e.g.~\cite{Taub54})
\begin{subequations}\label{displacements}
	\begin{align}
		\jau{a}{\mu} &= \Jau{a}{\mu} + \frac{1}{2}\nabla_\nu \bigl(\Jau{a}{\nu}\xiau{a}{\mu} - \Jau{a}{\mu}\xiau{a}{\nu}\bigr) + \calO\left(\xi^2\right)\,,\\
		\jbarau{a}{\mu} &= \Jau{a}{\mu} - \frac{1}{2}\nabla_\nu \bigl(\Jau{a}{\nu}\xiau{a}{\mu} - \Jau{a}{\mu}\xiau{a}{\nu}\bigr) + \calO\left(\xi^2\right)\,.
	\end{align}
\end{subequations}
We define the antisymmetric polarization tensor from the difference of the microscopic currents~\eqref{displacements} (since they carry opposite YM charges) as\footnote{Defining $j^\mu = \sum j_a^\mu t_a$ and $\overline{j}^\mu = \sum \overline{j}_a^\mu t_a$ the YM current~\eqref{Jmu} reads $\mathcal{J}^\mu = j^\mu - \overline{j}^\mu - \frac{c}{\ell} \bigl[ K_\nu, \Pi^{\mu\nu}\bigr]$.}
\begin{subequations}\label{defPia}
	\begin{align}
		\jau{a}{\mu}-\jbarau{a}{\mu}
		&= - c \nabla_\nu\Piau{a}{\mu\nu} + \calO\left(\xi^2\right)\,,\\
		\text{with}\quad\Piau{a}{\mu\nu} &\equiv \frac{1}{c}\Bigl(\Jau{a}{\mu}\xiau{a}{\nu} - \Jau{a}{\nu}\xiau{a}{\mu}\Bigr)\,.
	\end{align}
\end{subequations}
We shall extensively use the projection of the dipole moment perpendicular to the four-velocity of the particle, i.e. 
\begin{align}\label{defperp}
	\xiperpau{a}{\mu} = \perpa{a}{\mu}{\nu}\xiau{a}{\nu}\quad\text{where}\quad\perpa{a}{\mu}{\nu} &= \delta^{\mu}_{\nu} + \uau{a}{\mu}\ual{a}{\nu}\,.
\end{align}
This projection will constitute the basic variable of the present model (in the same way as in the previous DDM model~\cite{BL08, BL09}). We note that the polarization tensor can be written directly with the perpendicular projection of the dipole moment as
\begin{subequations}\label{Piperp}
	\begin{align}
		\Piau{a}{\mu\nu} \equiv \rhoa{a}\Bigl(\uau{a}{\mu}\xiperpau{a}{\nu} - \uau{a}{\nu}\xiperpau{a}{\mu}\Bigr)\,,
	\end{align}
	and that we have
	\begin{align}
		\ual{a}{\nu}\Piau{a}{\mu\nu} = \rhoa{a}\xiperpau{a}{\mu}\,.
	\end{align}
\end{subequations}
In the following we shall introduce for the three species of particles the ratio between their YM charge and their (gravitational type) mass, denoted $\eta_a$, so that the polarization tensor which couples to the YM field in Eqs.~\eqref{LYM} is actually given as
\begin{align}\label{Pihat}
	\Pihatau{a}{\mu\nu} \equiv \etaa{a}\Piau{a}{\mu\nu} \,,\qquad\text{(and also}~
	\Jihatau{a}{\mu} \equiv \etaa{a}\Jiau{a}{\mu}~\text{)}\,.
\end{align}
The coupling to ordinary gravity will be ensured by the polarization tensor~\eqref{defPia} itself; see further discussion around Eq.~\eqref{defeta}.

\section{Newtonian model for dipolar dark matter}
\label{sec:Newtonian}

In this section we present a physical model for dipolar dark matter in the Newtonian (non-relativistic) approximation, and we show that generic solutions of the model reproduce the MOND phenomenology. In Sec.~\ref{sec:covariant} we shall define a fully relativistic model.

\subsection{Ansätze for the non-relativistic approximation}

Our first task is to specify how the components of the YM field behave in the non-relativistic approximation (formally $c\to+\infty$). We assume the post-Newtonian (PN) ansatz
\begin{subequations}\label{PNansatz}
	\begin{align}
		\Kal{a}{0} &= \frac{1}{c^2}\phia{a} + \calO\left(c^{-4}\right)\,,\\
		\Kal{a}{i} &= \frac{1}{c^3}\Aal{a}{i} + \calO\left(c^{-5}\right)\,,
	\end{align}
\end{subequations}
where $\phi_a$ and $A_a^i$ have a finite non-zero limit when $c\to+\infty$ and the standard PN remainder is denoted $\calO(c^{-n})$. Then, we have to specify the way the constants $\ell$ in Eq.~\eqref{covder} and $\alpha$ in~\eqref{LYM} will scale with $c$. We find that in order to obtain non trivial solutions in the Newtonian limit, where the non-Abelian character of the YM field is preserved, and in particular where the cubic term in the YM action~\eqref{LYM} plays a role in the Newtonian limit, the constant $\ell$ must scale as $\sim c^{-3}$ while $\alpha$ must scale like $c^2$. Hence we introduce some constants $\bar{\ell}$ and $\bar{\alpha}$ supposed to be independent from $c$ and given by 
\begin{align}\label{kdef}
	\bar{\ell} = \ell c^3\,,\qquad\qquad \bar{\alpha} = \frac{\alpha}{c^2}\,.
\end{align}
Working out the limit $c\to+\infty$ while keeping $\bar{\ell}$ and $\bar{\alpha}$ constant, we find that the model admits a consistent non-relativistic approximation. In this limit the ``electric'' components of the YM field strength dominate, while the ``magnetic'' components are subdominant:
\begin{subequations}\label{PNansatzH}
	\begin{align}
		\Hau{a}{0i} &= \frac{1}{c^2} \Eal{a}{i} + \calO\left(c^{-4}\right)\,,\\
		\Hau{a}{ij} &= \calO\left(c^{-3}\right)\,.
	\end{align}
\end{subequations}
The electric YM field reads explicitly
\begin{align}\label{YMelectric}
	\Eal{a}{i} = \partial_i\phia{a} + \,\frac{1}{\bar{\ell}} \sum_{b,c} \epsilon_{abc} \Aal{b}{i}\phia{c}\,,
\end{align}
and the YM Lagrangian~\eqref{LYM} reduces to the purely electric expression 
\begin{align}\label{LYMnewtonian}
	L_\text{YM} = \sum_a \Biggl\{ \etaa{a} \Pial{a}{i} \Eal{a}{i} + \frac{1}{8\pi G} \biggl[ \Eal{a}{i} \Eal{a}{i} + \frac{\bar{\alpha}}{2}\sum_{b,c} \epsilon_{abc}\,\epsilon_{ijk}\Eal{a}{i}\Eal{b}{j}\Eal{c}{k} + \calO(\bar{\alpha}^2) \biggr]\Biggr\} + \calO(c^{-2})\,.
\end{align}
Note that we can ignore the cosmological constant in the Newtonian limit. This is justified by the fact that $\Lambda\sim\alpha^{-2}$ and $\alpha\sim c^2/a_0$ [see Eq.~\eqref{a0} below] hence we have $\Lambda=\calO(c^{-4})$. While the electric $\bm{E}$-field~\eqref{YMelectric} is given by the electric-type components of the ``bare'' field strength $H^{\mu\nu}$, see Eqs.~\eqref{PNansatz}, we can also define the analogue of the displacement $\bm{D}$-field (or induction field) which is given by the ``dressed'' field strength $Q^{\mu\nu}$ defined by Eq.~\eqref{exprQ}, and which can be seen as modified due to the DDM medium. Thus, we pose
\begin{subequations}\label{QPN}
	\begin{align}
		\Qau{a}{0i} &= \frac{1}{c^2} \Dal{a}{i} + \calO\left(c^{-4}\right)\,,\\
		\Qau{a}{ij} &= \calO\left(c^{-3}\right)\,,
	\end{align}
\end{subequations}
where the displacement field is related to the electric field by
\begin{align}\label{DaiPN}
	\Dal{a}{i} = \Eal{a}{i} + \frac{3\bar{\alpha}}{4} \sum_{b,c} \epsilon_{abc}\,\epsilon_{ijk}\Eal{b}{j}\Eal{c}{k} + \calO(\bar{\alpha}^2)\,.
\end{align}
Next, we specify our description of the particles. We have already described the three conserved currents~\eqref{defJa} that become in the non-relativistic approximation 
\begin{align}\label{defJa*}
	\Jstarau{a}{\mu} = \rhostara{a}\vau{a}{\mu}\,,\quad\text{such that}\quad\partial_\mu\!\Jstarau{a}{\mu} = 0\,,
\end{align}
where $\rho_a^*$ denotes the coordinate mass density and $v_a^\mu$ is the ordinary coordinate velocity (such that $v_a^0=c$ and $v_a^i = c \,u_a^i/u_a^0$). Concerning the dipole moment, we assume that the zero-th component behaves as $\xi_a^0=\calO(c)$ in the limit $c\to+\infty$, while the spatial components are finite, $\xi_a^i=\calO(1)$. This means that the perpendicular projection of the dipole moment defined by~\eqref{defperp} behaves as
\begin{subequations}\label{xiperpN}
	\begin{align}
		\xiperpau{a}{0} &= \calO(c^{-1})\,,\\
		\xiperpau{a}{i} &= \xiau{a}{i} - \frac{1}{c} \vau{a}{i}\xiau{a}{0} + \calO(c^{-2})\,.\label{xiperpNi}
	\end{align}
\end{subequations}
The spatial components of the dipole moment are non zero and finite in the Newtonian limit, and dominate over the zero-th components of the dipole which are negligible. As for the polarization tensor it admits a similar PN scaling 
%
\begin{subequations}\label{PolarN}
	\begin{align}
		\Piau{a}{0i} &= \Piau{a}{i} + \calO\left(c^{-2}\right)\,,\\
		\Piau{a}{ij} &= \calO\left(c^{-1}\right)\,,
	\end{align}
\end{subequations}
where $\Pi_a^{i}$ denotes the polarization vector in the Newtonian limit, given by
\begin{align}\label{defPiaN}
	\Piau{a}{i} = \rhostara{a}\xiperpau{a}{i}\,.
\end{align}

\subsection{Dipolar dark matter model in the Newtonian approximation}
\label{sec:DDMnewtonian}

Once we have specified the internal YM sector of the model~\eqref{LYMnewtonian}, it remains to add the dark matter particles. In our model (following Ref.~\cite{B07mond}) the dark matter medium will behave as a ``plasma'' made of positive and negative (gravitational-type) masses linked together by the internal force. In the Newtonian approximation the three YM conserved currents are given by~\eqref{defJa*}, and the polarization vector is defined by~\eqref{defPiaN}.

The main problem we face is adding the kinetic terms associated with the three species of particles and their dipole moments. We rely on a toy model discussed in Appendix~\ref{microscopic}, where three doublets of particles with $(m_\text{i},m_\text{g})=(m,\pm m)$ are coupled to the external gravitational field and the internal field. Furthermore, it is convenient (since we have this freedom) to add a parameter $\eta_a$ for the three species of particles defined as the ratio between their gravitational mass (or YM charge) when coupled to the YM field and their gravitational mass in the ordinary sense when coupled to the gravitational field. Thus, the inertial and gravitational masses of the three doublets of particles are 
\begin{subequations}\label{defeta}
	\begin{align}
		&\ma{a}{}_{\!\text{i}} \equiv \ma{a}\quad\text{with}\quad \ma{a}>0\,,\\
		&\ma{a}{}^{\!\text{grav}}_{\!\text{g}} = \pm \ma{a}\,,\quad\text{while}\quad
		\ma{a}{}^{\!\text{YM}}_{\!\text{g}} = \pm \etaa{a} \ma{a}\,.
	\end{align}
\end{subequations}

The looked-for kinetic terms as well as the interaction between the dipole moments and the fields to Newtonian order are given by Eq.~\eqref{toymodel} in Appendix~\ref{microscopic}. The kinetic term of the dipole moments is built from the Eulerian time derivative
\begin{align}\label{xipoint}
	\xidotperpau{a}{i} \equiv \frac{\dd \xi_{a\perp}^{i}}{\dd t} = \partial_t\xiperpau{a}{i} + \vau{a}{j} \partial_j\xiperpau{a}{i} + \calO\left(c^{-2}\right)\,.
\end{align}
Following~\eqref{toymodel}, the model will be valid up to quadratic order in the three dipole moments, thus with remainder $\calO(\xi^3_\perp)$. Finally, adding also the Newtonian gravitational field (with a non-relativistic Poisson potential $U$) and the baryons,\footnote{In the following we denote the quantities associated with the baryons with the subscript ``$\text{bar}$''.} we have the complete Lagrangian in the non-relativistic approximation: 
\begin{align}\label{LagN}
	L &= -\frac{1}{8\pi G} \,\partial_{i}U \partial_{i}U+ \rho_{*\text{bar}}\Bigl(U + \frac{v_\text{bar}^2}{2}\Bigr) + \sum_a \rhostara{a} \Bigl( \vau{a}{2}+ \frac{1}{4} \xidotperpau{a}{i}\!\xidotperpau{a}{i} \Bigr) \nn\\& + \sum_a \biggl\{ \Pial{a}{i} \left( \partial_{i}U + \etaa{a}\Eal{a}{i}\right) + \frac{1}{8\pi G}\biggl[ \Eal{a}{i} \Eal{a}{i} +  \frac{\bar{\alpha}}{2} \sum_{b,c}\epsilon_{abc} \,\epsilon_{ijk} \Eal{a}{i} \Eal{b}{j} \Eal{c}{k} \biggr]\biggr\} + \calO\left(\xi^3_\perp\right)\,.
\end{align}
For simplicity, we do not write the higher-order EFT remainder $\calO(\bar{\alpha}^2)$ nor the obvious PN remainder $\calO(c^{-2})$. 

We draw a particular attention to the terms in this Lagrangian that are specifically due to our hypothesis of gravitational dipoles, made of positive and negative gravitational-type masses, see~\eqref{defeta}. First, the term $\Pi_a^i \partial_{i}U$ is the gravitational analogue of the coupling between the electric dipole and the electric field $\bm{\Pi}_e\cdot\bm{E}$ in the Lagrangian of dielectrics (see the discussion in the Introduction). This term cannot exist in a fully covariant theory, as it implies a breakdown of the diffeomorphism invariance of GR. The second term in~\eqref{LagN} sharing the same breakdown is the kinetic term of the dipolar particles, i.e. $\rho_a^*\,v_a^2$, which is alone without the potential part built with $U$ (compare it with the baryon's contribution). However in Sec.~\ref{sec:covariant} we prove that it is possible to make such terms fully covariant at the price of introducing an extra degree of freedom in the form of the Khronon scalar field. Hence we propose a relativistic covariant Lagrangian in Eq.~\eqref{action} below.

To summarize, the dynamical variables of the Newtonian model~\eqref{LagN} are: (i) the gravitational Poisson potential $U$; (ii) the baryon's mass density $\rho^*_{\text{bar}}$ and velocity field $v_\text{bar}^i$ (or equivalently the conserved current $J_{*\text{bar}}^\mu$ such that $\partial_\mu J_{*\text{bar}}^\mu = 0$); (iii) the three species of dark matter particles described by the three conserved currents $J_{a*}^{\mu}$; (iv) the three dipole moments $\xi_{a\perp}^i$; (v) the YM fields $\phi_a$ and $A_a^i$ in Eq.~\eqref{PNansatz}.

The equation for the gravitational potential is just the Poisson equation including the contributions of the polarization vectors,
\begin{align}\label{eqU}
	\Delta U = - 4\pi G \Bigl( \rho_{*\text{bar}} - \sum_{a}\partial_i\!\Pial{a}{i}\Bigr)  + \calO\left(\xi^3_\perp\right) \,.
\end{align}
This equation is echoing Eq.~\eqref{rhoDM} from the MOND dielectric analogy. The equation of motion of the baryons is standard,
\begin{align}\label{EDMstandard}
	\frac{\dd v^i_\text{bar}}{\dd t} &= g_i \,,
\end{align}
where $g_i=\partial_i U$. However, in contrast to baryons the dipolar dark matter particles are accelerated not by the gravitational field but by its gradient or tidal field~\cite{B07mond}; in addition they are also accelerated by the gradient of the YM electric fields, thus
\begin{align}\label{eom1}
	\frac{\dd \vau{a}{\!i}}{\dd t} &= \frac{1}{2} \xiperpau{a}{j} \partial_i\bigl( g_j + \etaa{a}\Eal{a}{j}\bigr) + \calO\left(\xi^3_\perp\right) \,.
\end{align}
Recall that the YM electric field includes a crucial non-Abelian feature through the second term of Eq.~\eqref{YMelectric}. The evolution equations of the dipole moments read then\footnote{We have verified that one can vary the Lagrangian~\eqref{LagN} either with respect to the four components of the dipole moment $\xi_a^0$, $\xi_a^i$, or with respect to the three components of the perpendicular moment $\xi_{a\perp}^i$, making use of the relation~\eqref{xiperpNi}; the resulting equation~\eqref{evol1} is the same.}
\begin{align}\label{evol1}
	\xiddotperpau{a}{\,i} &= 2\bigl( g_i + \etaa{a}\Eal{a}{i}\bigr) + \calO\left(\xi^3_\perp\right) \,.
\end{align}
An important point about this equation is the remainder which is indicated as $\calO(\xi^3_\perp)$. In principle, since the remainder comes from the variation of a term of order $\calO(\xi^3_\perp)$ in the Lagrangian [see Eq.~\eqref{LDMinfinite}] we could expect a dependency in $\calO(\xi^2_\perp)$. However, it will also involve the double gradient of $g_i + \eta_a E_{ai}$, i.e. be of the type $\propto \xi_{a\perp}^j\xi_{a\perp}^k\partial_j\partial_k(g_i+\eta_aE_{ai})$. Now, solving~\eqref{evol1} in perturbation, we find that $g_i + \eta_a E_{ai}$ will itself be of the order $\calO(\xi_\perp)$, and therefore the remainder is actually of order $\calO(\xi^3_\perp)$.

Finally, it remains the equations for the YM fields, obtained by varying the non-relativistic Lagrangian with respect to $\phi_a$ and $A_a^i$. As we have already proved in the general relativistic model, the field equations can be nicely integrated in closed analytic form, with result given by Eq.~\eqref{integreq}. By taking the Newtonian limit of~\eqref{integreq}, or equivalently by varying~\eqref{LagN} with respect to $\phi_a$ and $A_{ai}$, we see that the induction $\bm{D}$-field~\eqref{DaiPN} in our model equals the polarization: 
\begin{align}\label{DaiPi}
	\Dal{a}{i} = -4\pi G \etaa{a}\Pial{a}{i}\,.
\end{align}
This relation corresponds to Eq.~\eqref{Polarization} in the Newtonian model of the Introduction, and can be regarded as the constitutive relation of the DDM medium. It can be inverted to provide the relation between the electric field and the dipole moments as
\begin{align}\label{Eexplicit}
	\Eau{a}{i}= -\frac{\omega_a^2}{2\eta_a} \xiperpau{a}{\,i} - \frac{3 \bar{\alpha}}{16} \sum_{b,c} \epsilon_{abc} \epsilon_{ijk} \frac{\omega_b^2 \omega_c^2}{\eta_b\eta_c} \xiperpau{b}{\,j} \xiperpau{c}{\,k}\; + \calO(\xi^3_\perp) \,,   
\end{align}
where we have introduced, for each of the YM particles, the ``plasma'' frequency 
\begin{align}
	\label{plasmafr}
	\omegaa{a}{}^{\!2} = 8\pi G \etaa{a}{}^2 \!\rhostara{a} \,.
\end{align}
As we show in Sec.~\ref{sec:MONDsolution} this relation yields the MOND behavior, with the constant $\bar{\alpha}$ essentially identified as the inverse of the MOND acceleration scale $a_0$. 

The relation~\eqref{DaiPi} can be combined with the evolution equations of the dipole moments~\eqref{evol1} to give an harmonic oscillator, whose frequency is given by the plasma frequency~\eqref{plasmafr}. Indeed, consistently with the order of approximation of the evolution equation~\eqref{evol1}, we obtain the equation obeyed by the dipole moments as
\begin{align}
	\label{Eq_dipole_order2}
	\xiddotperpau{a}{\,i} + \omegaa{a}{}^{\!2} \xiperpau{a}{\,i} = 2 g_i - \frac{3 \bar{\alpha}}{8} \sum_{b,c} \epsilon_{abc} \frac{\eta_a}{\eta_b\eta_c}\omegaa{b}{}^{\!2}\!\omegaa{c}{}^{\!2}\epsilon_{ijk}\xiperpau{b}{\,j} \xiperpau{c}{\,k} + \calO\left(\xi^3_\perp\right)\,.
\end{align}
The harmonic oscillator is forced by the gravitational field $g_i$, and by the second-order term on the right side, which we now prove to be responsible for the deep MOND limit of the model.\footnote{If we assume that $\xi$ has to emerge at a quantum level as a consequence of pair creation the natural order of magnitude for the dipole moment is the Compton wavelength $\xi\sim\lambda_\text{C}= \frac{h}{m c}$, thus the polarization is $\Pi\sim\frac{n h}{c}$ where $n$ is the number density of particles. Considering the dipole moments at equilibrium, $\Pi\sim\frac{g}{4\pi G}$ where $g$ is of the order of $a_{0}$ in the MOND regime, we can obtain an order of magnitude for the density of the dark matter particles, $n\sim\frac{c a_{0}}{4\pi h G} \sim 10^{35} \,\text{cm}^{-3}$. However, the whole setup of the present model is classical.}

\subsection{Solution reproducing the MOND phenomenology}
\label{sec:MONDsolution}

In this section, we show that generic solutions of the previous equations~\eqref{eqU}--\eqref{DaiPi} reproduce the phenomenology of MOND. The equation of the dipole moment has been obtained in~\eqref{Eq_dipole_order2}, where the ``plasma'' frequency is given by~\eqref{plasmafr}. We first note that from the equation of motion~\eqref{eom1} the velocity of the dipolar particles is a small term of order $\calO(\xi^2_\perp)$, where we used the fact already alluded to that $g_i + \eta_a E_{ai}$ is itself a small quantity of order $\calO(\xi_\perp)$. Therefore, from the continuity equation satisfied by the coordinate mass density $\rho_a^*$ [see Eq.~\eqref{defJa*}] we deduce that $\partial_t\rho_a^* = \calO(\xi^2_\perp)$ which implies that the time dependence of the plasma frequency~\eqref{plasmafr} contributes a small correction $\calO(\xi^3_\perp)$ in Eq.~\eqref{Eq_dipole_order2} and can be neglected. Furthermore, we note that up to negligible $\calO(\xi^3_\perp)$ terms the convective time derivative~\eqref{xipoint} reduces to the partial time derivative.

Thus, the equation of the dipole moment~\eqref{Eq_dipole_order2} appears as an ordinary (forced) harmonic oscillator with constant frequency and can be solved iteratively for small $\xi_{\perp}$. In first approximation the general solution reads
\begin{align}
	\bmxiperpa{a} = \frac{2 \bm{g}}{\omegaa{a}^2} + \bmXa{a} \cos{(\omegaa{a} t)} + \bmYa{a} \sin{(\omegaa{a} t)} + \calO\left(\xi^2_\perp\right)\,,
\end{align}
where the vectors $\bm{X}_{a}=(X_{a}^{i})$ and $\bm{Y}_{a}=(Y_{a}^{i})$ are time independent coefficients and can be seen as set by initial conditions for the evolution of the dipole. Iterating this solution to the next order we get
\begin{align}
	\label{xiSolution}
	\bmxiperpa{a} &= \frac{2 \bm{g}}{\omegaa{a}^2} + \bmXa{a} \cos{(\omegaa{a} t)} + \bmYa{a} \sin{(\omegaa{a} t)} \nn \\ & - \frac{3 \bar{\alpha}}{16}\sum_{b,c} \frac{\eta_{a}}{\eta_{b} \eta_{c}} \epsilon_{abc} \bigg[ \frac{8\omega_{c}^2}{\omega_{a}^2-\omega_{c}^2}\,\bm{g}\times\Bigl(\bmXa{c}\cos{(\omegaa{c} t)} + \bmYa{c} \sin{(\omegaa{c} t)}\Bigr) \nn \\
	&\quad+ \frac{\omega_{b}^2 \omega_{c}^2}{\omegaa{a}^2-{(\omegaa{b} + \omegaa{c})}^{2}}\bigg(\Bigl(\bmXa{b}\times\bmXa{c} - \bmYa{b}\times\bmYa{c}\Bigr)\cos{\bigl[(\omegaa{b} + \omegaa{c}) t\bigr]} + 2 \bmXa{b}\times\bmYa{c}\sin{\bigl[(\omegaa{b} + \omegaa{c}) t\bigr]} \bigg)  \nn  \\
	&\quad  + \frac{\omega_{b}^2 \omega_{c}^2}{\omega_{a}^2-{(\omega_{b} - \omega_{c})}^{2}} \bigg( 2 \bmXa{b}\times\bmYa{c}\sin{\bigl[(\omegaa{b} - \omegaa{c}) t\bigr]} + \Bigl(\bmXa{b}\times\bmXa{c} + \bmYa{b}\times\bmYa{c}\Bigr)\cos{\bigl[(\omegaa{b} - \omegaa{c}) t\bigr]} \bigg)\bigg] \nn \\ &+ \calO\left(\xi^3_\perp\right)\,.
\end{align}
The modified Poisson equation~\eqref{eqU} is then
\begin{subequations}\label{modPoisson}
	\begin{align}
		\bm{\nabla}\cdot\bm{\mathcal{D}} = - 4\pi G \rho_{*\text{bar}} \,,
	\end{align}
	where $\bm{\mathcal{D}} \equiv \bm{g} - 4\pi G \sum_a \bm{\Pi}_{a}$ explicitly reads 
	\begin{align}\label{Delta}
		\bm{\mathcal{D}} &= \bm{g}\Bigl(1-\sum_{a} \frac{1}{\eta_{a}^2}\Bigr) - \sum_{a} \frac{\omega_{a}^2}{2 \eta_{a}^2} \Bigl( \bmXa{a} \cos{(\omegaa{a} t)} + \bmYa{a} \sin{(\omegaa{a} t)}\Bigr)  \nn \\ & + \frac{3 \bar{\alpha}}{4}\sum_{a,b,c} \frac{\epsilon_{abc}}{\eta_{a} \eta_{b} \eta_{c}}\frac{\omega_{a}^2 \omega_{c}^2}{\omega_{a}^2-\omega_{c}^2} \,\bm{g}\times\Bigl(\bmXa{c}\cos{(\omegaa{c} t)} + \bmYa{c} \sin{(\omegaa{c} t)}\Bigr) \nn\\ 
		& + \frac{3 \bar{\alpha}}{32} \sum_{a,b,c} \frac{\omega_{a}^2 \omega_{b}^2 \omega_{c}^2 \epsilon_{abc} }{\eta_{a} \eta_{b} \eta_{c} } \bigg[\frac{1}{\omegaa{a}^2-{(\omegaa{b} + \omegaa{c})}^{2}}\bigg(\Bigl(\bmXa{b}\times\bmXa{c} - \bmYa{b}\times\bmYa{c}\Bigr)\cos{\bigl[(\omegaa{b} + \omegaa{c}) t\bigr]} \nn \\ 
		&+ 2 \bmXa{b}\times\bmYa{c}\sin{\bigl[(\omegaa{b} + \omegaa{c}) t\bigr]} \bigg)  + \frac{1}{\omega_{a}^2-{(\omega_{b} - \omega_{c})}^{2}} \bigg( 2 \bmXa{b}\times\bmYa{c}\sin{\bigl[(\omegaa{b} - \omegaa{c}) t\bigr]} \nn \\ & + \Bigl(\bmXa{b}\times\bmXa{c} + \bmYa{b}\times\bmYa{c}\Bigr)\cos{\bigl[(\omegaa{b} - \omegaa{c}) t\bigr]} \bigg) \bigg] \,.
	\end{align}
\end{subequations}
We now show that there are generic solutions that reproduce the MOND phenomenology in the physical case where
\begin{enumerate}
	\item The three YM mass ratios of the particles satisfy
	\begin{align}\label{relationeta}
		\sum_{a} \frac{1}{\eta_{a}^2} \simeq 1\,.
	\end{align}
	\item There are at least two kinds of dark matter particles whose plasma frequencies are comparable, and we choose without loss of generality
	\begin{align}\label{assume1}
		\Delta_\omega \equiv \vert\omegaa{2}-\omegaa{3}\vert ~\ll~ \omegaa{2}\,,\,\omegaa{3}\,.
	\end{align}
\end{enumerate}
In this situation the particle 1 plays a special role and for instance, we could have $\eta_1\simeq 1$ and $\eta_2, \eta_3 \gg 1$. The MOND effect will be due to the polarization associated with the dipole moment of the particle 1. Actually, for simplicity, we can further assume the hierarchy of frequencies
\begin{equation}\label{assume2}
	{(\omegaa{2}-\omegaa{3})}^2~\ll~ {\omegaa{1}}^2 ~\ll~ {(\omegaa{2}+\omegaa{3})}^2\, .
\end{equation}

\subsubsection{Generic stationary solution}

Searching for a stationary solution to the set of equations of motion, the requirement that the evolution equation on dark matter pairs~\eqref{eom1} should be time independent for the particular solution implies that the vectors $\bm{X}_{a}$ and $\bm{Y}_{a}$ are subject to the constraints
\begin{align}\label{relation_X_eq_dm}
	\bmXa{a}\cdot\,\bm{g} = 0\,, \qquad \bmYa{a}\cdot\,\bm{g} = 0\,, \qquad \bmXa{a}\cdot\bmXa{a} = \bmYa{a}\cdot\bmYa{a}\,, \qquad \bmXa{a}\cdot\bmYa{a} = 0 \,.
\end{align}
Next, we assume without loss of generality that the gravitational field $\bm{g}=\bm{\nabla} U$ is in the $x$-direction of the Cartesian coordinate system $\{x,y,z\}$; thus $g=U'(x)$ where the prime denotes the derivative with respect to the $x$-coordinate. The constraints~\eqref{relation_X_eq_dm} give then $X_1^x=Y_1^x=0$, and imply that $ \bmXa{2} \times \bmXa{3} = \pm \bmYa{2} \times \bmYa{3}$. In order to have a solution consistent with the presence of baryons in the modified Poisson equation~\eqref{eqU} and compatible with the plasma frequencies~\eqref{assume2} we shall need to restrict ourselves to the case
\begin{align}\label{suplcond}
	\bmXa{2} \times \bmXa{3} = \bmYa{2} \times \bmYa{3}\,.
\end{align}
Furthermore, we consider the unidimensional problem where $\bm{X}_{a}$ and $\bm{Y}_{a}$ are only functions of the $x$-coordinate through the gravitational field $g(x)$. A simple solution satisfying all the requirements assumes $\bm{X}_1=\bm{Y}_1=\bm{0}$,\footnote{This condition is also compatible with a baryon distribution which is stationary.} remembering that 1 is the privileged particle which will give the MOND effect. Thus, combining all the previous constraints, including the hierarchy~\eqref{assume2} for the plasma frequencies, we obtain
\begin{align}
	\bm{\mathcal{D}}= \frac{3 \bar{\alpha}}{8  \eta_{1} \eta_{2} \eta_{3}}  \,\omega_{2}^2  \omega_{3}^2 \biggl(\bmXa{2} \times \bmXa{3} \cos{\bigl[(\omegaa{2} - \omegaa{3}) t\bigr]} + 2 \bigl(\bmXa{2} \times \bmYa{3}+\bmXa{3} \times \bmYa{2}\bigr) \sin{\bigl[(\omegaa{2} - \omegaa{3}) t\bigr]}  \biggr) \,.
\end{align}
Using the condition~\eqref{suplcond} the most general solution for $\bm{X}_{2,3}$ and $\bm{Y}_{2,3}$ is
\begin{align}
	\label{explicit_solution}
	\bmXa{2} = X_2\begin{bmatrix}
		0 \\[0.2cm]
		\cos\theta_2 \\[0.2cm]
		\sin\theta_2
	\end{bmatrix},
	~\bmYa{2} = \epsilon X_2\begin{bmatrix}
		0 \\[0.2cm]
		-\sin\theta_2\\[0.2cm]
		\cos\theta_2
	\end{bmatrix},
	~\bmXa{3} = X_3\begin{bmatrix}
		0 \\[0.2cm]
		\cos\theta_3\\[0.2cm]
		\sin\theta_3
	\end{bmatrix},
	~\bmYa{3} = \epsilon X_3\begin{bmatrix}
		0 \\[0.2cm]
		-\sin\theta_3 \\[0.2cm]
		\cos\theta_3 
	\end{bmatrix},
\end{align}
where $\epsilon = \pm 1 $, and $\theta_2$, $\theta_3$ are constant angles. Moreover, the norms of $\bmXa{a}$ and $\bmYa{a}$ necessarily scale as $\frac{g}{\omega_{a}^2}$ according to equation~\eqref{xiSolution}, so we shall have
\begin{align}\label{explicitDelta}
	X_2 = k_2\,\frac{g}{\omega_2^2}\,, \qquad X_3 = k_3\,\frac{g}{\omega_3^2}\,,
\end{align}
with some numerical factors $k_2$ and $k_3$. For the latter solution the modified Poisson equation~\eqref{modPoisson} becomes, as long as we consider sufficiently short time $t\ll\frac{1}{\Delta_{\omega}}$, see~\eqref{assume1}:
\begin{align}\label{deepMOND}
	\bigl({\mathcal{D}}^x\bigr)' = \Bigl(-\frac{3\bar{\alpha}}{8}\frac{k_2 k_3}{\eta_1\eta_2\eta_3}\sin\theta_{23} \,g^2\Bigr)' =  - 4\pi G \rho_{*\text{bar}} \,,
\end{align}
with $\theta_{23}=\theta_{2}-\theta_{3}$, and where the $g^{2}$ dependency originally comes from the cubic term in the action~\eqref{LYM}. Therefore, the equation~\eqref{deepMOND} gives the deep MOND limit with the acceleration scale
\begin{align}\label{expressiona0}
	a_0 = - \frac{8}{3\bar{\alpha}}\frac{\eta_1\eta_2\eta_3}{k_2 k_3 \sin\theta_{23}} \,,
\end{align}
where we recall that $\alpha=\bar{\alpha}c^2$ is the EFT scale introduced in the action~\eqref{LYM}. Once the modified Poisson equation is solved the baryons obey the standard dynamics, see~\eqref{EDMstandard}. 

Besides the EFT scale there are some dimensionless numerical factors in~\eqref{expressiona0}. While the $\eta_a$'s are properties of the particles and have their full place in~\eqref{expressiona0}, the factors $k_2$, $k_3$ and $\sin\theta_{23}$ are a priori questionable because they are expected to depend on initial conditions at the beginning of the formation of the system (a given galaxy). We do not control these numerical factors, but in principle their values should be fixed by the dynamics of the system that led to this stationary solution. We conjecture that they should be more or less universal, depending only on the scenario of formation of the system, and that this scenario should be essentially alike for most galaxies; more work should be done to clarify this point. If all the dimensionless numerical factors are of the order unity the MOND constant is approximately
\begin{align}\label{a0}
	a_0 \simeq \frac{c^2}{\alpha}\,,
\end{align}
which fixes the scale of the perturbative EFT expansion in Eq.~\eqref{LYM}. For the latter solution, we can determine the three polarization vectors $\bm{\Pi}_a$ and successively [using~\eqref{DaiPi} and the inverse of~\eqref{DaiPN}] the $\bm{D}_a$ and $\bm{E}_a$ fields, see Eq.~\eqref{Eexplicit}. At this stage, an important test of the consistency of our solution is to find the expression for the YM fields $\phi_a$ and $\bm{A}_a$ related to the electric field $\bm{E}_{a}$ by Eq.~\eqref{YMelectric}. With the explicit expression~\eqref{explicit_solution}, we can determine a solution such that (up to second order in perturbation)
\begin{subequations}\label{phiasol}
	\begin{align}
		\phia{1} &= -\frac{U}{\eta_{1}}+ \delta\!\phia{1}\,,\\
		\phia{2} &= \delta\!\phia{2}\,,\\
		\phia{3} &= \delta\!\phia{3}\,,
	\end{align}
\end{subequations}
where $U$ is the Newtonian potential ($g=U'$) and the perturbation $\delta\phi_a$ depends on $t$ and $x$. For simplicity we keep our hierarchy of frequencies, i.e. we perform appropriate limits following~\eqref{assume2}. We find that the perturbation $\delta\phi_1$ is determined by an integral over the perturbation in the electric field $\delta E_{1}^x$ and over explicitly determined $\delta\phi_2$, $\delta\phi_3$:
\begin{subequations}\label{phiasol2}
	\begin{align}
		\delta\!\phia{1} &= \int \Big(\delta \Eau{1}{x} + \frac{\eta_{1} g}{U \eta_{2} \eta_{3}}\big(\etaa{3} \delta\!\phia{2}+\etaa{2} \delta\!\phia{3}\big)\Big) \dd x \,,  \\
		\delta\!\phia{2} &= \frac{3 \bar{\alpha} U g \epsilon}{ 2 \eta_{1} \eta_{3} \sin{[(\omega_{2}-\omega_{3})t +\epsilon \theta_{23} ]}} \Big(\frac{k_{3}}{k_{2}} - \cos{[(\omega_{2}-\omega_{3})t +\epsilon \theta_{23} ]} \Big)\,, \\
		\delta\!\phia{3} &=  \frac{3 \bar{\alpha} U g \epsilon}{ 2 \eta_{1} \eta_{2} \sin{[(\omega_{2}-\omega_{3})t +\epsilon \theta_{23} ]}} \Big(\frac{k_{2}}{k_{3}} - \cos{[(\omega_{2}-\omega_{3})t +\epsilon \theta_{23} ]} \Big)\,.
	\end{align}
	We have, in more details,
	\begin{align}
		\delta\!\phia{1} &= \frac{3\bar{\alpha} \epsilon}{2 \eta_{2} \eta_{3}} \int \bigg[\frac{1}{\sin[ (\omega_{2}-\omega_{3})t + \epsilon \theta_{23}]} \Big( \frac{k_{2}}{k_{3}}+\frac{k_{3}}{k_{2}} + \frac{k_{2} k_{3}}{4} \sin^{2}{[(\omega_{2}-\omega_{3})t+\epsilon \theta_{23}]} \nn \\
		&~-2 \cos{[(\omega_{2}-\omega_{3})t+\epsilon \theta_{23}]} \Big)  + \frac{k_{2} k_{3} }{4} \sin[(\omega_2-\omega_{3})t - \epsilon \theta_{23}] \bigg] g^{2}(x) \dd{x}\, .
	\end{align}
\end{subequations}
To satisfy Eq.~\eqref{YMelectric} we can choose the associated YM potential vectors as\footnote{In our explicit solution, the YM potential vectors are proportional to $\bar{\ell}$. When inserted into the expression of the YM electric fields~\eqref{YMelectric}, we see that the $\bar{\ell}$ dependency will cancel out. This explains why the MOND acceleration scale~\eqref{expressiona0} is finally independent from $\bar{\ell}$.}
\begin{subequations}\label{Aasol2}
	\begin{align}
		\bmAa{1} &= \bm{0}\,,\\
		\bmAa{2} &= \frac{\bar{\ell} \,\eta_{1} }{U} \Big[ \bmEa{3} \Bigl(1+\frac{\eta_{1}}{U}\delta\!\phia{1}\Bigr) - \delta\!\phia{3}{}^{\!'} \bm{x}  \Big] \,, \\
		\bmAa{3} &= - \frac{\bar{\ell} \,\eta_{1} }{U} \Big[ \bmEa{2} \Bigl(1+\frac{\eta_{1}}{U}\delta\!\phia{1}\Bigr) - \delta\!\phia{2}{}^{\!'} \bm{x}\Big] \,.
	\end{align}
\end{subequations}
Finally, we recall that the solution describes the deep MOND limit. Higher-order corrections are in principle given by the EFT theory, which may even provide a full non-perturbative expression (using the theory of effective Lagrangians~\cite{Dunne05,ItzyksonZuber}) which would ideally describe some kind of transition between the deep MOND regime and the high-acceleration weak-field Newtonian regime (or GR). However, such a transition to the Newtonian regime is not described in the present model and is left for future work.

\subsubsection{Generic averaged solution}

Alternatively to the previous derivation, the MOND solution can also be recovered by some suitable average over time, assuming the hierarchy of plasma frequencies~\eqref{assume2}. For instance, one may remove the high frequency ``noise'' associated with the high frequencies $\omega_2$ and $\omega_3$ by performing a time average of the solution~\eqref{xiSolution} over the short period $T_2=2\pi/\omega_2$ or $T_3=2\pi/\omega_3$. In this case, with the same assumptions as before (notably $\bm{X}_1=\bm{Y}_1=\bm{0}$), we recover the same solution with MOND scale~\eqref{expressiona0}. A way to recover the MOND solution without any assumption is to average over a time interval $T_\text{obs}$ corresponding to some observation of dark matter effects, provided that the associated frequency $\omega_\text{obs}=2 \pi/T_\text{obs}$ fits into the plasma frequencies~\eqref{assume2} as
\begin{equation}\label{assumeobs}
	{(\omegaa{2}-\omegaa{3})}^2~\ll~ \omega_\text{obs}^2 ~\ll~ {\omegaa{1}}^2 ~\ll~ {(\omegaa{2}+\omegaa{3})}^2\,.
\end{equation}
In this case the average over the observed time $T_\text{obs}$, i.e. $\langle f(t)\rangle_\text{obs}=\frac{1}{T_\text{obs}}\int_0^{T_\text{obs}}f(t)\dd t$, gives for the $\bm{D}$-field
\begin{subequations}
	\begin{align}
		\langle\bmDal{1}\rangle_\text{obs} &= -\frac{\bm{g}}{\eta_{1}}+\frac{3 \bar{\alpha}}{16 \eta_{2} \eta_{3}} \Bigl(\bmxa{2} \times \bmxa{3} + \bmya{2} \times \bmya{3} \Bigr)\,, \\
		\langle\bmDal{2}\rangle_\text{obs} &= -\frac{\bm{g}}{\eta_{2}} - \frac{3 \bar{\alpha}}{16 \pi \eta_{1} \eta_{3}} \frac{\omega_\text{obs}}{\omega_{2} -\omega_{3}}\biggl( \sin\bigl(\pi \tfrac{\omega_{2}+\omega_{3}}{\omega_\text{obs}}\bigr) \,\bm{g} \times \bmxa{3} + \Bigl[1- \cos\bigl(\pi \tfrac{\omega_{2}+\omega_{3}}{\omega_\text{obs}}\bigr) \Bigr]\,\bm{g} \times \bmya{3} \biggr)\,, \\
		\langle\bmDal{3}\rangle_\text{obs} &= -\frac{\bm{g}}{\eta_{3}} - \frac{3 \bar{\alpha}}{16 \pi \eta_{1} \eta_{2}} \frac{\omega_\text{obs}}{\omega_{2} -\omega_{3}} \biggl( \sin\bigl(\pi \tfrac{\omega_{2}+\omega_{3}}{\omega_\text{obs}}\bigr) \,\bm{g} \times \bmxa{2} + \Bigl[1- \cos\bigl(\pi \tfrac{\omega_{2}+\omega_{3}}{\omega_\text{obs}}\bigr)\Bigr] \,\bm{g} \times \bmya{2} \biggr)\,,
	\end{align}
\end{subequations}
where we have defined $\bm{x}_{a}\equiv\bm{X}_{a} \omega_{a}^2$ and $\bm{y}_{a}\equiv\bm{X}_{a} \omega_{a}^2$. The components of $\langle\bmDal{2}\rangle_\text{obs}$ and $\langle\bmDal{3}\rangle_\text{obs}$ proportional to $\bm{g}\times\bm{x}_a$ and $\bm{g}\times\bm{y}_a$ are not observable as they are orthogonal to $\bm{g}$ (since $\bm{g}$ depends only on the $x$-coordinate). Thus, the modified Poisson equation factor reads at leading order [after use of Eq.~\eqref{relationeta}]
\begin{align}
	\langle\bm{\mathcal{D}}^x\rangle = \frac{3 \bar{\alpha}}{16  \eta_{1} \eta_{2} \eta_{3}}  \Bigl(\bmxa{2} \times \bmxa{3} + \bmya{2} \times \bmya{3} \Bigr)^x \,,
\end{align}	
which, again, generically provides the MOND formula as the norm of each $\bm{x}_{a}$ and $\bm{y}_{a}$ is proportional to $g$. 	

\section{Relativistic model for dipolar dark matter}
\label{sec:covariant}

As already mentioned, the Newtonian model based on the action~\eqref{LagN} cannot be the Newtonian limit of a full covariant model in GR. Of course many terms are straightforwardly written in covariant form, for instance, the Newtonian kinetic term $\propto\partial_{i}U \partial_{i}U$ gives the curvature scalar $R$, the baryon contribution is just given by the scalar mass density $\rho_\text{bar}$ such that $J_\text{bar}^\mu=c \rho_\text{bar}u_\text{bar}^\mu$ is covariantly conserved, $\nabla_\mu J_\text{bar}^\mu=0$, and we have already written in Eq.~\eqref{LYM} the YM action in full covariant form (also containing the cosmological constant). But, for instance, the gravitational field $g_i=\partial_i U$ in the action, which appears through the dipolar coupling $\Pi_a^i g_i$, cannot a priori be made covariant. 

Nevertheless, the model~\eqref{LagN} can be ``covariantized'' by the introduction of the Khronon scalar field, which is the Stückelberg field associated with the broken diffeomorphism invariance~\cite{BM11,Sand11,BS24}. The Khronon field $\tau$ defines a preferred spatial foliation through the hypersurface orthogonal unit-timelike vector field 
\begin{subequations}\label{nmu_def}
	\begin{align}
		n_\mu &= - \frac{c}{\Qcal} \nabla_\mu \tau\,,\\ 
		\text{with}~~\Qcal &\equiv c\sqrt{- g^{\mu\nu} \nabla_\mu \tau \nabla_\nu \tau}\,,
	\end{align}
\end{subequations}
such that $g_{\mu\nu} n^\mu n^\mu  = -1$ and $n^0 > 0$. The covariant spacelike acceleration (chosen to have the dimension of an ordinary acceleration) is
\begin{align}\label{amu_def}
	a_\mu =  c^2 n^\nu \nabla_\nu n_\mu = - c^2 \gamma_{\mu}^{\nu}  \,\nabla_\nu \ln \Qcal \,,
\end{align}
where $\gamma_{\mu}^{\nu} \equiv \delta_\mu^{\nu}  + n_\mu n^\nu$ is the projector onto the foliation, such that $n^\mu a_\mu = 0$.\footnote{In the following we shall be careful at not confusing $\gamma_{\mu}^{\nu}$ with the perpendicular projection with respect to the 4-velocity, $\perp_{a\mu}^{\nu}=\delta_\mu^{\nu}  + u_{a\mu} u_a^\nu$.} The four-acceleration~\eqref{amu_def} reduces to the usual Newtonian acceleration in the Newtonian limit, so it can be used to covariantize the dipolar term $\Pi_a^i g_i$. More precisely, using the covariant expression of the polarization tensor~\eqref{Piperp}, we readily compute the Newtonian limit in adapted coordinates $\tau=t$ (so-called unitary gauge):
\begin{align}\label{covariant1}
	\Piau{a}{\mu\nu} \!\ual{a}{\mu} \,a_\nu = \Piau{a}{i} g_i + \calO\left(c^{-2}\right)\,.
\end{align}
Therefore, the left side of~\eqref{covariant1} is a good candidate for a covariant term in the action. The other term we have to treat in the Newtonian action is the kinetic term $v_a^2$ that appears alone, i.e. without the potential term $U$. For this term, we compute
\begin{align}\label{covariant2}
	-2\rhoa{a} c^2\left(1+\uau{a}{\mu} n_\mu\right) = \rhostara{a} \vau{a}{2} + \calO\left(c^{-2}\right)\,,
\end{align}
where $n_\mu$ is given by~\eqref{nmu_def}, so that again, the left side is a good candidate for the covariant form. 

However, there is a caveat here: to perform the Newtonian limits~\eqref{covariant1} and~\eqref{covariant2} which reproduce the required terms in the Newtonian model, we have to assume the standard PN behavior of the metric components,
\begin{equation}\label{PNhypoth}
	g_{00}=-1+\calO(c^{-2}) \,, \qquad g_{0i}=\calO(c^{-3})\,, \qquad g_{ij}=\delta_{ij}+\calO(c^{-2}) \,,
\end{equation}
conjointly with the use of the unitary gauge; in particular, we assume that the $0i$ component of the metric is a small PN term of order $\calO(c^{-3})$. But we know from~\cite{Flanagan23, BS24} that the PN hypothesis~\eqref{PNhypoth} is incompatible with the unitary gauge for the Khronon field. Indeed the Khronon will admit a PN expansion 
\begin{align}\label{PNkhronon}
	\tau=t+\frac{\sigma}{c^2} + \calO(c^{-4})\,,
\end{align}
where $\sigma$ is a function of space and time, which implies in principle $g_{0i}=\calO(c^{-1})$ when going to the unitary gauge, instead of $\calO(c^{-3})$ (see~\cite{Flanagan23, BS24} for further discussion). Therefore, there will be a price to be paid for using Eqs.~\eqref{covariant1}--\eqref{covariant2} with the unitary gauge, namely the Newtonian model~\eqref{LagN} and the MOND limit will be recovered but only for stationary systems. 

The general-relativistic action of the model, $S=\int \dd^{4}x \sqrt{-g}\,L$, is therefore defined by 
\begin{align}\label{action}
	L &= \frac{c^4}{16 \pi G }\biggl[ R-2\Lambda + \sum_a \biggl( - \Hal{a}{\mu\nu} \Hau{a}{\mu\nu} + \frac{\alpha}{2} \sum_{b,c} \epsilon_{abc} \Hal{a}{\mu\tau}\Haul{b}{\tau}{\nu}\Hduala{c}{\mu\nu} + \calO(\alpha^2) \biggr) \biggr]- \rho_\text{bar} c^2\nn\\
	&~+ \sum_a\biggl[\Piau{a}{\mu\nu}\Bigl( \ual{a}{\mu} a_\nu-\frac{c^2}{2}\etaa{a}\Hal{a}{\mu\nu}\Bigr)-2\rhoa{a} \left( c^{2} (1+\uau{a}{\mu} n_\mu)-\frac{1}{8}\xidotperpau{a}{\lambda}\xidotperpal{a}{\lambda}\right) + \mathcal{O}\left(\xi_\perp^3\right)\biggr]\,,
\end{align}
where $R$ is the curvature scalar, the $\text{SU}(2)$ YM field $H_a^{\mu\nu}$ and its dual ${H}_a^{\star\mu\nu} = \varepsilon^{\mu\nu\rho\sigma}{H}_{a\rho\sigma}$ are described in Sec.~\ref{sec:YM}, the polarization tensor $\Pi_{a}^{\mu\nu}$ is defined by Eqs.~\eqref{Piperp} in terms of three currents $J_a^\mu = c \rho_a u_a^\mu$ and the perpendicular components of the dipole moment~\eqref{defperp}. All space-time indices are raised and lowered with the space-time metric $g_{\mu\nu}$ and the time derivatives of the dipole moment are defined by
\begin{align}
	\xidotperpau{a}{\lambda} &\equiv c \uau{a}{\mu}\nabla_\mu\xiperpau{a}{\lambda}\,,
	\qquad \xiddotperpau{a}{\lambda} \equiv c \uau{a}{\mu}\nabla_\mu\xidotperpau{a}{\lambda}\,,\quad\textit{etc.}\,,
\end{align}
which reduce to e.g.~\eqref{xipoint} in the Newtonian limit. In the action~\eqref{action} $\rho_\text{bar}$ denotes the conserved scalar mass density of the baryons. Since we are interested in the dark matter at the scale of galaxies, it is sufficient to model the ordinary matter just by baryons (stars and gas in galaxies). However, we have in mind that the baryons actually represent the standard model of particle physics.

The only dependence on the Khronon field in Eq.~\eqref{action} is through the unit vector $n_\mu$ in the last term and the acceleration $a_\nu$ in the penultimate term, see~\eqref{nmu_def}--\eqref{amu_def}. The cosmological constant was introduced in Eq.~\eqref{LYM} and is naturally of the order of magnitude of the inverse square of the EFT constant $\alpha$, see~\eqref{Lambda}, which is itself related to the MOND acceleration scale by Eq.~\eqref{a0}. Hence, a remarkable feature of the model is that the cosmological constant should be of the same order of magnitude as the square of the MOND acceleration,
\begin{align}
	\Lambda \sim \frac{a_0^2}{c^4}\,,
\end{align}
in good agreement with cosmological observations.\footnote{The related numerical coincidence that $H_0 \sim a_0/c$ was pointed out by Milgrom~\cite{Milg1, Milg2, Milg3}.}

We vary the relativistic action with respect to all the particles and dynamical fields. First, the equation for the baryons naturally leads to the geodesic equation
\begin{align}
	\dot{u}_\text{bar}^\mu \equiv c \, u_\text{bar}^\nu\!\nabla_\nu u_\text{bar}^\mu=0\,.
\end{align}
The equation of motion of the YM particles is more involved, notably because we have to vary the projection operator in the definition of the perpendicular projection of the dipole moment~\eqref{defperp}. We obtain\footnote{It is convenient to perform a ``fluid'' variation~\cite{Taub54, Ca91, CaKh92} with respect to the conserved current $J_a^\mu$ (constrained to satisfy $\nabla_\mu J_a^\mu=0$), namely
	\begin{align*}
		\Jau{a}{\nu}\left[ \nabla_\nu\left(\frac{\delta L}{\delta J_a^{\mu}}\right) - \nabla_\mu\left(\frac{\delta L}{\delta J_a^{\nu}}\right) \right] = 0\,.
\end{align*}}
\begin{align}\label{eoma}
	\pdotal{a}{\mu} &= 2 \uau{a}{\nu} \bigl(a_{\nu} n_{\mu} - a_{\mu} n_{\nu}\bigr)  - \xiperpau{a}{\lambda}\bigl(\nabla_{\mu} a_{\lambda}+  c^{2} \etaa{a} \uau{a}{\nu} \nabla_{\mu} \Hal{a}{\nu\lambda}\bigr) + \frac{c}{2} \uau{a}{\nu} R_{\lambda\sigma \mu \nu} \xiperpau{a}{\sigma} \xidotperpau{a}{\lambda} + \calO\left(\xi_\perp^3\right)\,,
\end{align}
where $R_{\lambda\sigma \mu \nu}$ is the Riemann tensor and we have posed
\begin{align}\label{pamu}
	c\pal{a}{\mu} &\equiv \ual{a}{\mu}\Bigl(2 c^{2} + \xiperpau{a}{\lambda} a_\lambda + \frac{1}{4}\xidotperpau{a}{\lambda}\xidotperpal{a}{\lambda}\Bigr) - \xiperpal{a}{\mu}  \uau{a}{\lambda} \Bigl(a_{\lambda}+ \frac{1}{2}\xiddotperpal{a}{\lambda}\Bigr) - c^{2} \etaa{a} \xiperpau{a}{\lambda}\Hal{a}{\mu\lambda} + \calO\left(\xi_\perp^3\right)\,.
\end{align}
Then, the equation obtained by variation of the dipole moment itself (i.e. $\xi_a^\mu$) reads
\begin{align}\label{eqxi}
	\perpa{a}{\mu}{\nu} \Bigl(\xiddotperpau{a}{\nu}+2a^\nu\Bigr) = 2c^2\! \etaa{a}\uau{a}{\nu}\!\Haul{a}{\mu}{\nu} + \calO\left(\xi_\perp^3\right)\,,
\end{align}
while the equations for the $\text{SU}(2)$ YM fields have already been obtained in~\eqref{fieldeqYM}. We have seen that they can be integrated in the most general case with the result 
\begin{align}\label{integreqa}
	\Qau{a}{\mu\nu} = - \frac{4\pi G}{c^2}\,\etaa{a}\Piau{a}{\mu\nu}\,,
\end{align}
the $Q_a^{\mu\nu}$'s being given by~\eqref{exprQ}. 

It remains the equations associated to the two ``gravitational'' fields, namely the metric $g_{\mu\nu}$ and the Khronon $\tau$. By varying with respect to the Khronon we obtain the covariant conservation law $\nabla_{\mu}\mathcal{S}^{\mu}=0$ where the current vector $\mathcal{S}^\mu$ reads
\begin{align}\label{conscurrent}
	\mathcal{S}^{\mu}= \sum_a \frac{1}{\Qcal}\biggl[ n^{\mu} \nabla_{\nu}\bigl(\rhoa{a} \gamma_{\lambda}^{\nu} \xiperpau{a}{\lambda} \bigr) - 2 \rhoa{a} \gamma_{\nu}^{\mu} \uau{a}{\nu}  + \rhoa{a}  \xiperpau{a}{\nu}\bigl(\gamma_{\nu}^{\mu} n^{\lambda}\nabla_{\lambda} \ln{\Qcal} - \frac{a^{\mu} n_{\nu}}{c^2} \bigr) \biggr] \,,
\end{align}
where $\Qcal$ is defined by~\eqref{nmu_def} and $\gamma_{\mu}^{\nu} = \delta_\mu^{\nu}  + n_\mu n^\nu$. Finally the variation with respect to the metric yields the Einstein field equations\footnote{For notational convenience we transfer the cosmological constant on the left side of the equations although it really belongs to the dark matter sector of our model, see Eq.~\eqref{LYM}.}
\begin{align}\label{EFE2}
	G^{\mu\nu} + \Lambda g^{\mu\nu} = \frac{8 \pi G}{c^4} \Bigl[T_\text{bar}^{\mu\nu} + \Tcal^{\mu\nu}\Bigr]\,,
\end{align}
where $T_\text{bar}^{\mu\nu}$ is the baryon's stress-energy tensor, and $\Tcal^{\mu\nu}$ is interpreted as the contribution of the ``dark matter'' and reads after some simplification with Eq.~\eqref{eqxi}, 
\begin{align}\label{Tdmmunu}
	\Tcal^{\mu\nu} &= \sum_a\Biggl\{  \rhoa{a}\biggl[ \uau{a}{\mu}\uau{a}{\nu}\Bigl(2 c^{2} + \xiperpau{a}{\lambda} a_\lambda + \frac{1}{4}\xidotperpau{a}{\lambda}\xidotperpal{a}{\lambda}\Bigr) + 2 c^{2} \uau{a}{\lambda} n_\lambda n^\mu n^\nu + 2 \xiperpau{a}{\lambda}\gamma_\lambda^{(\mu} a^{\nu)} + \frac{1}{2} \xiddotperpau{a}{(\mu}\xiperpau{a}{\nu)} \nn\\
	&\quad\quad - 2 c^{2} \etaa{a} \uau{a}{\lambda}\xiperpau{a}{(\mu}\Haul{a}{\nu)}{\lambda}\biggr] - \frac{c}{2} \nabla_\lambda\Bigl(\rhoa{a}\uau{a}{(\mu}\bigl[ \xidotperpau{a}{\nu)}\xiperpau{a}{\lambda}-\xiperpau{a}{\nu)}\xidotperpau{a}{\lambda}\bigr]\Bigr) - c^{2} n^\mu n^\nu \nabla_\lambda\bigl(\rhoa{a}\gamma_\sigma^\lambda \xiperpau{a}{\sigma}\bigr) + \calO\left(\xi_\perp^3\right)\nn\\
	&\quad +\frac{c^4}{4\pi G} \biggl[\Haul{a}{\mu}{\sigma} \Hau{a}{\nu\sigma} - \frac{1}{4}g^{\mu\nu} \Hal{a}{\lambda\sigma} \Hau{a}{\lambda\sigma} + \frac{\alpha}{4} \sum_{b,c} \epsilon_{abc} \Haul{a}{\mu}{\lambda} \Haul{b}{\nu}{\sigma} \Hduala{c}{\lambda\sigma} + \calO(\alpha^2)\biggr]\Biggr\} \,.
\end{align}
A useful alternative form of the DM stress-energy tensor is obtained by using the expression of the particle's ``momentum''~\eqref{pamu} and the expression of the Khronon's conserved current~\eqref{conscurrent}. After further reduction, we obtain the following expression
\begin{align}\label{Tdmmunu2}
	\Tcal^{\mu\nu} =& - c^2 \Qcal \,n^\mu \mathcal{S}^\nu 
	\nn\\
	& + \sum_a\Biggl\{\rhoa{a}\biggl[ c \uau{a}{\nu}\bigl(\pau{a}{\mu} - 2 c n^\mu\bigr) - c \xiperpau{a}{\lambda} \gamma^\nu_\lambda \,\nabla^\mu\ln\Qcal + \frac{1}{4} \Bigl( \xiddotperpau{a}{\mu}\xiperpau{a}{\nu}-\xiddotperpau{a}{\nu}\xiperpau{a}{\mu}\Bigr)\biggr]
	\nn\\
	&\quad 
	- \frac{c}{2} \nabla_\lambda\Bigl(\rhoa{a}\uau{a}{(\mu}\bigl[ \xidotperpau{a}{\nu)}\xiperpau{a}{\lambda}-\xiperpau{a}{\nu)}\xidotperpau{a}{\lambda}\bigr]\Bigr) + \calO\left(\xi_\perp^3\right)\nn\\
	&\quad +\frac{c^4}{4\pi G} \biggl[ - \frac{1}{4}g^{\mu\nu} \Hal{a}{\lambda\sigma} \Hau{a}{\lambda\sigma} + \frac{\alpha}{2} \sum_{b,c} \epsilon_{abc} \varepsilon^{\nu\rho\sigma(\tau}\Haul{a}{\lambda)}{\rho} \Hal{b}{\sigma\tau} \Haul{c}{\mu}{\lambda} + \calO(\alpha^2)\biggr]\Biggr\}\,.
\end{align}
We have verified by a long calculation starting from~\eqref{Tdmmunu2} that the stress-energy tensor is conserved as a consequence of the matter and field equations:
\begin{align}\label{divTdm}
	\nabla_\nu\Tcal^{\mu\nu} = 0\,.
\end{align}

It is straightforward to check that the non relativistic limits of the above covariant equations of motion in the unitary gauge for the Khronon are the same as the equations for the Newtonian model, see Sec.~\ref{sec:DDMnewtonian}. However the covariant formulation presents the additional equation~\eqref{conscurrent} corresponding to the Khronon field. The non relativistic limit of this equation takes the form of a continuity equation and reads 
\begin{align}
	\sum_a \Bigl[\partial_{t}\!\rhotaua{a} + \partial_i\bigl(\rhotaua{a}\,v_\tau^i\bigr)\Bigr] = \calO(c^{-2})\,,
\end{align}
where we pose [recalling Eq.~\eqref{PNkhronon}]
\begin{align}
	\rhotaua{a} \equiv 2 \rhostara{a} + \partial_{i}\bigl(\rhostara{a} \xiperpau{a}{i}\bigr)\,,\qquad
	v_\tau^i \equiv -\partial_{i}\sigma\,.
\end{align}
Thus, if we assume that the system is stationary, i.e. $\sum_{a}  \partial_{t}\rho_{a\tau}=0$, the Khronon field can be taken equal to the coordinate time, $\tau=t$, and we are allowed to take the Newtonian limit using the PN ansatz~\eqref{PNhypoth} in the unitary gauge, so we recover the same behavior as the Newtonian model of Sec.~\ref{sec:DDMnewtonian} in the non-relativistic limit. This restriction to stationary systems is acceptable because the MOND phenomenology (rotation curves of galaxies, Tully-Fisher relation, etc.) concerns mostly stationary systems~\cite{FamMcG12}.

We also point out that the restriction to stationary systems is because we have insisted that we should covariantize the Newtonian model~\eqref{LagN} by introducing the extra Khronon degree of freedom. If we accept that in the low acceleration regime there is a breakdown of the local Lorentz invariance, we could probably build a relativistic extension of the model using only $3$+$1$ variables, which retrieves the Newtonian model~\eqref{LagN} in the non-relativistic approximation; in this case the MOND limit would be valid even for non-stationary systems. 

\section{Conclusion}
\label{sec:conclusion}

We have presented an effective theory for MOND based on a dipolar dark matter (DDM) medium whose internal dynamics is ruled by a non-Abelian Yang-Mills vector field based on the gauge symmetry group $\text{SU}(2)$. This field is supposed to represent a new sector of the standard model of particle physics, currently containing three layers associated to the $\text{U}(1)$, $\text{SU}(2)$ and $\text{SU}(3)$ gauge symmetries. The new interaction is coupled to gravity (we coin it a ``graviphoton'') and emerges in the low acceleration regime, below the MOND acceleration scale $a_0$, where Newtonian dynamics and general relativity have not been verified to any accuracy. Using arguments from the effective field theory (EFT) we show that it permits to recover the deep MOND regime at the scale of galaxies in a natural way, without having to introduce by hand an ad hoc function in the action. 

The model involves a breakdown of Lorentz invariance (and general covariance) in the low acceleration regime. Hence, the general covariance would appear as an approximate symmetry only valid in the high acceleration limit. Nevertheless, we show that the general covariance can be restored even for low accelerations, at the price of introducing a supplementary degree of freedom in the form of the Khronon scalar field, whose gradient is time-like. In the fully relativistic (``covariantized'') version of the model the deep MOND regime is still recovered but we find a restriction to systems being stationary.

The main attractive feature of the model, we believe, is that the MOND phenomenology is connected to some deeper underlying Physics --- a new sector of the particle physics model. In its present form, the model has some incompleteness:
\begin{enumerate}
	\item The deep MOND regime is recovered but remains disconnected from the usual regime of Newtonian gravity or GR. In ordinary MOND theories, there is a function which plays the role of the interpolating function of MOND and permits to study the high acceleration regime of the theory. In this model the transition to the high acceleration regime is missing but may be implemented by using the theory of effective Lagrangians~\cite{EulerHeisenberg, Dunne05, ItzyksonZuber}. On the other hand, we know that the MOND interpolating function is highly constrained by observations in the Solar System~\cite{Milg09, BN11, HFAG15, VNT24}, so it is perhaps a good point that the transition to the high acceleration regime in the present model looks more complicated than a simple interpolating function.
	\item In the present model, there is no mechanism to support the assumptions regarding the dipole properties~\eqref{relationeta} and~\eqref{assume2} that we use to obtain the MOND phenomenology. It would be interesting to associate these assumptions with a concrete particle physics model. Such a model could also provide constraints on the particle properties, as well as on the nature of the particles involved (fermions or bosons).
	\item The model does not explain the effects commonly attributed to dark matter in cosmology, which occur at large scales in cosmological perturbations. However, we expect that it could be extended to mimic the dark matter in cosmology by adding a kinetic term in the action for the Khronon scalar field, and showing that the Khronon equations can be recast into the framework of the generalized dark matter (GDM) model~\cite{Hu98, KST16} (as was done for the Aether-Scalar-Tensor theory~\cite{SZ21} and the Khronon theory~\cite{BS24}).
	\item The MOND solutions we have found in Sec.~\ref{sec:MONDsolution} depend on details about initial conditions at the start of the formation of the system. Numerical simulations should be performed in order to evaluate the universality of such initial conditions and the impact on the measured value of the MOND acceleration scale.
\end{enumerate}
More work should be done to improve the model in these directions. 

\acknowledgments
We would like to thank Gilles Esposito-Far\`ese for interesting discussions and for the careful reading of this manuscript.

\appendix

\section{Newtonian microscopic model for gravitational dipoles}
\label{microscopic}

We consider dark matter composed of doublets of sub-particles, one with positive gravitational mass $m_\text{g} = m$ and one with negative gravitational mass $m_\text{g} = - m$, where $m_\text{i} = m$ represents their always positive inertial mass. The ordinary particle $(m_\text{i}, m_\text{g}) = (m, m)$ will always be attracted by an external mass made of ordinary matter; however, the exotic particle $(m_\text{i}, m_\text{g}) = (m, -m)$ will be repelled by the same external mass. In addition, the two sub-particles will repel each other (see Ref.~\cite{B07mond} for a discussion). We consider a Newtonian toy model in which the particles interact with the Newtonian gravitational potential $U$ and with some internal potential $\Phi$.\footnote{For simplicity we consider only one type of particles and a single scalar potential $\Phi$ for the internal field, instead of the YM machinery considered in the main text. However, we introduce a parameter $\eta$ to remind our distinction~\eqref{defeta} between gravitational mass and internal charge.} The non relativistic Lagrangian for the positive and negative gravitational-mass particles writes 
\begin{align}
	L_\text{toy} &= \!\sum_{(m_\text{i}, m_\text{g}) = (m, m)} m\biggl[ \frac{1}{2}\bm{v}^{2}+U(\bm{y})+ \eta \Phi(\bm{y}) \biggr] \nn\\ 
	&+ \!\sum_{(m_\text{i}, m_\text{g}) = (m, -m)}\!m\biggl[ \frac{1}{2}\overline{\bm{v}}^{2}-U(\overline{\bm{y}})- \eta \Phi(\overline{\bm{y}}) \biggr] 
	\,,
\end{align}
where $\bm{v}=\dd\bm{y}/\dd t$ and $\overline{\bm{v}}=\dd\overline{\bm{y}}/\dd t$. The interactions between these particles create microscopic dipole moments. We define the dipole moment $\bm{\xi}$ as the separation between the two types of particles, and the position of the dipolar particle as the center of inertial mass $\bm{x}$:  
\begin{align}\label{dipolevar}
	\bm{\xi} \equiv \bm{y} - \overline{\bm{y}}\,, \qquad
	\bm{x} \equiv \frac{\bm{y} + \overline{\bm{y}}}{2}\,.
\end{align}
Expanding the Lagrangian to any order with respect to the dipole variables~\eqref{dipolevar} we obtain the formal infinite series
\begin{equation}\label{LDMinfinite}
	L_\text{toy} = \sum_\text{dipoles} m \biggl[ \bm{v}^{2} + \frac{1}{4}\dtc{\bm{\xi}} + \sum_{p=0}^{+ \infty}\frac{\xi^{2p+1}}{2^{2p}(2p+1)!}\partial_{2p+1}\bigl(U +  \eta \Phi\bigr)(\bm{x}) \biggr]\,,
\end{equation}
where $\xi^{2p+1}\partial_{2p+1}$ is a shortcut for $\xi^{i_{1}}\xi^{i_{2}}\cdots\xi^{i_{2p+1}}\partial_{i_{1}}\partial_{i_{2}}\cdots\partial_{i_{2p+1}}$, with $\partial_i\equiv\partial/\partial x^i$ (and implicit summations over the $2p+1$ spatial indices). Varying with respect to $\bm{x}$ and $\bm{\xi}$ we obtain the equations of motion of the dipolar particles and the evolution equations of the dipoles:
\begin{subequations}
	\begin{align}
		\frac{\dd v^{i}}{\dd t} &= \sum_{p=0}^{+ \infty} \frac{\xi^{2p+1}}{2^{2p+1}(2p+1)!}\partial_{i}\partial_{2p+1}\bigl(U + \eta \Phi\bigr)\,,\\
		\frac{\dd^{2}\xi^{i}}{\dd t^{2}} &= \sum_{p=0}^{+ \infty} \frac{\xi^{2p}}{2^{2p-1}(2p)!}\partial_{i}\partial_{2p}\bigl(U +  \eta \Phi\bigr)\,.
	\end{align}
\end{subequations}
To keep the relevant physics we restrict ourselves at the Lagrangian level to the second order in the dipole moment. This choice is motivated as it provides simple though non trivial equations of motion, given by
\begin{subequations}
	\begin{align}
		\frac{\dd v^{i}}{\dd t} &= \frac{1}{2}\xi^{j}\partial_{i}\partial_{j}\bigl(U + \eta \Phi\bigr)+ \mathcal{O}(\xi^{3})\,,\\
		\frac{\dd^{2}\xi^{i}}{\dd t^{2}} &= 2\partial_{i}\bigl(U +  \eta \Phi\bigr)+ \mathcal{O}(\xi^{2})\,.
	\end{align}
\end{subequations}
The Lagrangian of gravitational dipoles to this order takes the form
\begin{equation}\label{toymodel}
	L_\text{toy}=\sum_\text{dipoles} m \biggl[ \bm{v}^{2} + \frac{1}{4}\dtc{\bm{\xi}} + \bm{\xi}\cdot\bm{\nabla}\bigl(U + \eta \Phi\bigr) + \mathcal{O}(\xi^{3})\biggr]\,.
\end{equation}
This form provides the motivation for the kinetic terms of the dipolar particles and the interaction terms with $U$ and the internal field that we have adopted in the DDM model~\eqref{LagN}.

\bibliography{ListRef_DDM24}

@Article{BDgef11,
  Title                    = {Improving relativistic MOND with Galileon k-mouflage},
  Author                   = {Babichev, E. and Deffayet, C. and Esposito-Far{\`e}se, G.},
  Journal                  = {Phys. Rev. D},
  Year                     = {2011},
  Pages                    = {061502(R)},
  Volume                   = {84},

  Eprint                   = {arXiv:1106.2538 [gr-qc]},
  Slaccitation             = {%%CITATION = GR-QC 0311052;%%}
}

@Article{Bek04,
  Title                    = {Relativistic gravitation theory for the modified
 {N}ewtonian dynamics paradigm},
  Author                   = {Bekenstein, J.D.},
  Journal                  = {Phys. Rev. D},
  Year                     = {2004},
  Pages                    = {083509},
  Volume                   = {70},

  Eprint                   = {astro-ph/0403694}
}

@Article{BekM84,
  Title                    = {Does the missing mass problem signal the
 breakdown of {N}ewtonian gravity?},
  Author                   = {Bekenstein, J.D. and Milgrom, M.},
  Journal                  = {Astrophys. J.},
  Year                     = {1984},
  Pages                    = {7},
  Volume                   = {286}
}

@Article{BK15,
  Title                    = {Theory of Dark Matter Superfluidity},
  Author                   = {Berezhiani, L. and Khoury, J.},
  Journal                  = {Phys. Rev. D},
  Year                     = {2015},
  Pages                    = {103510},
  Volume                   = {92}
}

@misc{BK25review,
      title={Superfluid Dark Matter}, 
      author={Lasha Berezhiani and Giordano Cintia and Valerio De Luca and Justin Khoury},
      year={2025},
      eprint={2505.23900},
      archivePrefix={arXiv},
      primaryClass={astro-ph.CO},
      url={https://arxiv.org/abs/2505.23900}, 
}

@Article{BB14,
  Title                    = {Phenomenology of Dark Matter via a Bimetric Extension of General Relativity},
  Author                   = {Bernard, L. and Blanchet, L.},
  Journal                  = {Phys. Rev. D},
  Year                     = {2015},
  Pages                    = {103536},
  Volume                   = {91},
  Eprint                   = {arXiv:1410.7708 [astro-ph]}
}

@Article{B07mond,
  Title                    = {Gravitational polarization and the phenomenology of
 {MOND}},
  Author                   = {Blanchet, L.},
  Journal                  = {Class. Quant. Grav.},
  Year                     = {2007},
  Pages                    = {3529},
  Volume                   = {24},

  Eprint                   = {astro-ph/0605637}
}

@Article{Blanchet,
  Title                    = {A class of non-metric couplings to gravity},
  Author                   = {Blanchet, L.},
  Journal                  = {Phys. Rev. Lett.},
  Year                     = {1992},
  Pages                    = {559},
  Volume                   = {69}
}

@Article{BL09,
  Title                    = {Dipolar dark matter and dark energy},
  Author                   = {Blanchet, L. and Le Tiec, A.},
  Journal                  = {Phys. Rev. D},
  Year                     = {2009},
  Pages                    = {023524},
  Volume                   = {80},

  Eprint                   = {arXiv:0901.3114 [astro-ph]}
}

@Article{BL08,
  Title                    = {Model of dark matter and dark energy based on
 gravitational polarization},
  Author                   = {Blanchet, L. and Le Tiec, A.},
  Journal                  = {Phys. Rev. D},
  Year                     = {2008},
  Pages                    = {024031},
  Volume                   = {78},

  Eprint                   = {astro-ph/0804.3518}
}

@Article{BM11,
  Title                    = {Modified gravity approach based on a preferred
 time foliation},
  Author                   = {Blanchet, L. and Marsat, S.},
  Journal                  = {Phys. Rev. D},
  Year                     = {2011},
  Pages                    = {044056},
  Volume                   = {84},

  Eprint                   = {arXiv:1107.5264 [gr-qc]}
}

@Article{BN11,
  Title                    = {External field effect of modified {N}ewtonian dynamics
 in the Solar system},
  Author                   = {Blanchet, L. and Novak, J\'er\^ome},
  Journal                  = {Mon. Not. Roy. Astron. Soc.},
  Year                     = {2011},
  Pages                    = {2530},
  Volume                   = {412}
}

@Article{BGef07,
  Title                    = {Field-theoretical formulations of {MOND}-like gravity},
  Author                   = {J.-P.~Bruneton and G.~Esposito-Far\`ese},
  Journal                  = {Phys. Rev. D},
  Year                     = {2007},
  Pages                    = {124012},
  Volume                   = {76},

  Eprint                   = {arXiv:0705.4043 [gr-qc]}
}

@Article{Ca91,
  Title                    = {Convective variational approach to relativistic thermodynamics of dissipative fluids},
  Author                   = {B.~Carter},
  Journal                  = {Proc. R. Soc. Lond. A},
  Year                     = {1991},
  Pages                    = {45},
  Volume                   = {433}
}

@Article{CaKh92,
  Title                    = {Equivalence of convective and potential variational derivations of covariant superfluid dynamics},
  Author                   = {B.~Carter and I.~M.~Khalatnikov},
  Journal                  = {Phys. Rev. D},
  Year                     = {1992},
  Pages                    = {4536},
  Volume                   = {45}
}

@Article{DEW11,
  Title                    = {Nonlocal metric formulations of MOND with sufficient lensing},
  Author                   = {Deffayet, C. and Esposito-Far\`ese, G. and Woodard,
 R.},
  Journal                  = {Phys. Rev. D},
  Year                     = {2011},
  Pages                    = {124054},
  Volume                   = {84},

  Eprint                   = {arXiv:1106.4984 [gr-qc]}
}

@Article{FamMcG12,
  Title                    = {Modified {N}ewtonian dynamics ({MOND}):
 Observational phenomenology and relativistic
 extensions},
  Author                   = {B.~Famaey and S.~McGaugh},
  Journal                  = {Living Rev. Relativ.},
  Year                     = {2012},
  Pages                    = {10},
  Volume                   = {15},

  Eprint                   = {arXiv:1112.3960 [astro-ph.CO]}
}

@Article{GD92,
  Title                    = {Analysis of X-ray galaxy clusters in the framework
 of modified {N}ewtonian dynamics},
  Author                   = {Gerbal, D. and Durret, F. and Lachi\`eze-Rey, M. and
 Lima-Neto, G.},
  Journal                  = {Astron. Astrophys.},
  Year                     = {1992},
  Pages                    = {395},
  Volume                   = {262}
}

@Article{HZL08,
  Title                    = {A nonuniform dark energy fluid: Perturbation equations},
  Author                   = {A.~Halle and H.~S.~Zhao and B.~Li},
  Journal                  = {Astrophys. J. Suppl.},
  Year                     = {2008},
  Pages                    = {1},
  Volume                   = {177},

  Eprint                   = {arXiv:0711.0958 [astro-ph]}
}

@Book{ItzyksonZuber,
  Title                    = {Quantum Field Theory},
  Author                   = {Itzykson, C. and Zuber, J.-B.},
  Publisher                = {McGraw-Hill},
  Year                     = {1980},

  Address                  = {New-York}
}

@Article{LBMZ08,
  Title                    = {Testing Alternative Theories of Dark Matter with
 the CMB},
  Author                   = {B.~Li and J.~D.~Barrow and D.~F.~Mota and
 H.~S.~Zhao},
  Journal                  = {Phys. Rev. D},
  Year                     = {2008},
  Pages                    = {064021},
  Volume                   = {78},

  Eprint                   = {arXiv:0805.4400}
}

@Article{bimond1,
  Title                    = {Bimetric MOND gravity},
  Author                   = {Milgrom, M.},
  Journal                  = {Phys. Rev. D},
  Year                     = {2009},
  Pages                    = {123536},
  Volume                   = {80}
}

@Article{Milg09,
  Title                    = {MOND effects in the inner solar system},
  Author                   = {Milgrom, M.},
  Journal                  = {Mon. Not. Roy. Astron. Soc.},
  Year                     = {2009},
  Pages                    = {474},
  Volume                   = {399}
}

@Article{Milg1,
  Title                    = {A modification of the {N}ewtonian dynamics as a
 possible alternative to the hidden mass hypothesis},
  Author                   = {Milgrom, M.},
  Journal                  = {Astrophys. J.},
  Year                     = {1983},
  Pages                    = {365},
  Volume                   = {270}
}

@Article{Milg2,
  Title                    = {A modification of the {N}ewtonian dynamics:
 Implications for galaxies},
  Author                   = {Milgrom, M.},
  Journal                  = {Astrophys. J.},
  Year                     = {1983},
  Pages                    = {371},
  Volume                   = {270}
}

@Article{Milg3,
  Title                    = {A Modification of the {N}ewtonian dynamics:
 Implications for galaxy systems},
  Author                   = {Milgrom, M.},
  Journal                  = {Astrophys. J.},
  Year                     = {1983},
  Pages                    = {384},
  Volume                   = {270}
}

@Article{Sand05,
  Title                    = {A tensor-vector-scalar framework for modified dynamics 
 and cosmic dark matter},
  Author                   = {Sanders, R.H.},
  Journal                  = {Mon. Not. Roy. Astron. Soc.},
  Year                     = {2005},
  Pages                    = {459},
  Volume                   = {363},

  Eprint                   = {astro-ph/0502222}
}

@Article{Sand99,
  Title                    = {The virial discrepancy in clusters of galaxies
 in the context of modified {N}ewtonian dynamics},
  Author                   = {Sanders, R.H.},
  Journal                  = {Astrophys. J.},
  Year                     = {1999},
  Pages                    = {L23},
  Volume                   = {512}
}

@Article{Sand97,
  Title                    = {A stratified framework for scalar-tensor
 theories of modified dynamics},
  Author                   = {Sanders, R.H.},
  Journal                  = {Astrophys. J.},
  Year                     = {1997},
  Pages                    = {492},
  Volume                   = {480},

  Eprint                   = {astro-ph/9612099}
}

@Article{Sand11,
  Title                    = {Hiding Lorentz invariance violation with MOND},
  Author                   = {Sanders, R. H.},
  Journal                  = {Phys. Rev. D},
  Year                     = {2011},

  Month                    = {Oct},
  Pages                    = {084024},
  Volume                   = {84},

  Issue                    = {8},
  Numpages                 = {5},
  Publisher                = {American Physical Society}
}

@Article{Scherk,
  Title                    = {Antigravity: A crazy idea?},
  Author                   = {J. Scherk},
  Journal                  = {Physics Letters B},
  Year                     = {1979},
  Number                   = {3},
  Pages                    = {265 - 267},
  Volume                   = {88}
}

@Article{Sk08,
  Title                    = {Generalizing tensor-vector-scalar cosmology},
  Author                   = {Skordis, Constantinos},
  Journal                  = {Phys. Rev. D},
  Year                     = {2008},
  Pages                    = {123502},
  Volume                   = {77},

  Eprint                   = {arXiv:0801.1985 [astro-ph]},
  Issue                    = {12},
  Numpages                 = {10},
  Publisher                = {American Physical Society}
}

@Article{SMFB06,
  Title                    = {Large scale structure in {B}ekenstein's theory of
 relativistic modified {N}ewtonian dynamics},
  Author                   = {C.~Skordis and D.~F.~Mota and P.~ G.~Ferreira and
 C.~B{\oe}hm},
  Journal                  = {Phys. Rev. Lett.},
  Year                     = {2006},
  Pages                    = {011301},
  Volume                   = {96},

  Eprint                   = {arXiv:astro-ph/0505519}
}

@Article{Taub54,
  Title                    = {General relativistic variational principle for
 perfect fluids},
  Author                   = {A.~H.~Taub},
  Journal                  = {Phys. Rev.},
  Year                     = {1954},
  Pages                    = {1468},
  Volume                   = {94}
}

@Article{Zu10,
  Title                    = {Vector field models of modified gravity and the dark sector},
  Author                   = {Zuntz, J. and Zlosnik, T. G and Bourliot, F. and Ferreira,
 P. G. and Starkman, G. D.},
  Journal                  = {Phys. Rev. D},
  Year                     = {2010},

  Month                    = {May},
  Pages                    = {104015},
  Volume                   = {81},

  Eprint                   = {arXiv:1002.0849},
  Issue                    = {10},
  Numpages                 = {16},
  Publisher                = {American Physical Society}
}

@article{SZ21,
  title={New relativistic theory for modified Newtonian dynamics},
  author={Skordis, Constantinos and Z{\l}o{\'s}nik, Tom},
  journal={Physical review letters},
  volume={127},
  number={16},
  pages={161302},
  year={2021},
  publisher={APS}
}

@article{Flanagan23,
  title={Khronometric theories of modified Newtonian dynamics},
  author={Flanagan, {\'E}anna {\'E}},
  journal={The Astrophysical Journal},
  volume={958},
  number={2},
  pages={107},
  year={2023},
  publisher={IOP Publishing},
  eprint = "2302.14846",
  archivePrefix = "arXiv"
}

@article{Horava2009,
  title={Quantum gravity at a Lifshitz point},
  author={Ho{\v{r}}ava, Petr},
  journal={Physical Review D},
  volume={79},
  number={8},
  pages={084008},
  year={2009},
  publisher={APS},
  eprint = "0901.3775",
  archivePrefix = "arXiv"
}

@article{Blas2009,
  title={A healthy extension of Horava gravity},
  author={Blas, D and Pujolas, O and Sibiryakov, S},
  journal={Physical review letters},
  volume={104},
  number={1},
  pages={181302},
  year={2010},
  publisher={APS},
  eprint = "0909.3525",
  archivePrefix = "arXiv"
}

@article{Blas2011,
  title={Models of non-relativistic quantum gravity: The Good, the bad and the healthy},
  author={Blas, Diego and Pujolas, Oriol and Sibiryakov, Sergey},
  journal={Journal of High Energy Physics},
  volume={2011},
  number={4},
  pages={1--53},
  year={2011},
  publisher={Springer}
}

@article{ZFS07,
	author	= "T.~G.~Z{\l}o{\'s}nik and P.~G.~Ferreira and G.~D.~Starkman",
	journal	= "Phys. Rev. D",
	volume	= "75",
	pages	= "044017",
	year	= "2007",
	eprint	= "astro-ph/0607411"}

@article{BS24,
   title={Relativistic Khronon theory in agreement with modified Newtonian dynamics and large-scale cosmology},
   volume={2024},
   ISSN={1475-7516},
   url={http://dx.doi.org/10.1088/1475-7516/2024/11/040},
   DOI={10.1088/1475-7516/2024/11/040},
   number={11},
   journal={Journal of Cosmology and Astroparticle Physics},
   publisher={IOP Publishing},
   author={Blanchet, Luc and Skordis, Constantinos},
   year={2024},
   month=nov, 
   pages={040},
   eprint={2404.06584},
   archivePrefix={arXiv},
   primaryClass={gr-qc}
}

@article{Mendoza12,                                                                                                                                                                                                 
    author = "Mendoza, S. and Bernal, T. and Hidalgo, J. C. and Capozziello, S.",
    editor = "Beltran Jimenez, Jose and Ruiz Cembranos, Jose Alberto and Dobado, Antonio and Lopez Maroto, Antonio and De la Cruz Dombriz, A.",
    title = "{MOND as the weak-field limit of an extended metric theory of gravity}",
    eprint = "1202.3629",
    archivePrefix = "arXiv",
    primaryClass = "gr-qc",
    doi = "10.1063/1.4734465",
    journal = "AIP Conf. Proc.",
    volume = "1458", 
    number = "1",    
    pages = "483--486",
    year = "2012"
}

@article{Woodard14,  
        author         = "Woodard, R. P.",
        title          = "{Nonlocal metric realizations of MOND}",
        journal        = "Can. J. Phys.",
        volume         = "93",
        year           = "2015",
        number         = "2",
        pages          = "242-249",
        doi            = "10.1139/cjp-2014-0156",
        eprint         = "arXiv:1403.6763",
        archivePrefix  = "arXiv",
        primaryClass   = "astro-ph.CO",
        reportNumber   = "UFIFT-QG-14-02",
        SLACcitation   = "%%CITATION = ARXIV:1403.6763;%%"
}

@article{Milgrom19,
    author = "Milgrom, Mordehai",
    title = "{Noncovariance at low accelerations as a route to MOND}",
    eprint = "1908.01691",
    archivePrefix = "arXiv",
    primaryClass = "gr-qc",
    doi = "10.1103/PhysRevD.100.084039",
    journal = "Phys. Rev. D",
    volume = "100",
    number = "8",
    pages = "084039",
    year = "2019"
}

@article{DAmbrosio20,
   title={Non-linear extension of non-metricity scalar for MOND},
   volume={811},
   ISSN={0370-2693},
   url={http://dx.doi.org/10.1016/j.physletb.2020.135970},
   DOI={10.1016/j.physletb.2020.135970},
   journal={Physics Letters B},
   publisher={Elsevier BV},
   author={D’Ambrosio, Fabio and Garg, Mudit and Heisenberg, Lavinia},
   year={2020},
   month=dec, pages={135970} }

@article{HFAG15,
   title={Combined Solar system and rotation curve constraints on MOND},
   volume={455},
   ISSN={1365-2966},
   url={http://dx.doi.org/10.1093/mnras/stv2330},
   DOI={10.1093/mnras/stv2330},
   number={1},
   journal={Monthly Notices of the Royal Astronomical Society},
   publisher={Oxford University Press (OUP)},
   author={Hees, Aurélien and Famaey, Benoit and Angus, Garry W. and Gentile, Gianfranco},
   year={2015},
   month=nov, pages={449–461} 
}

@article{PS05,
   title={New constraints on modified Newtonian dynamics from galaxy clusters},
   volume={364},
   ISSN={1365-2966},
   url={http://dx.doi.org/10.1111/j.1365-2966.2005.09590.x},
   DOI={10.1111/j.1365-2966.2005.09590.x},
   number={2},
   journal={Monthly Notices of the Royal Astronomical Society},
   publisher={Oxford University Press (OUP)},
   author={Pointecouteau, Etienne and Silk, Joseph},
   year={2005},
   month=dec, pages={654–658} 
}

@misc{VNT24,
      title={Testing MOND on small bodies in the remote solar system}, 
      author={David Vokrouhlický and David Nesvorný and Scott Tremaine},
      year={2024},
      eprint={2403.09555},
      archivePrefix={arXiv},
      primaryClass={astro-ph.CO},
      url={https://arxiv.org/abs/2403.09555}, 
}

@article{BK16,
   title={Dark matter superfluidity and galactic dynamics},
   volume={753},
   ISSN={0370-2693},
   url={http://dx.doi.org/10.1016/j.physletb.2015.12.054},
   DOI={10.1016/j.physletb.2015.12.054},
   journal={Physics Letters B},
   publisher={Elsevier BV},
   author={Berezhiani, Lasha and Khoury, Justin},
   year={2016},
   month=feb, pages={639–643} 
}

@incollection{Dunne05,
  title={Heisenberg--Euler effective Lagrangians: basics and extensions},
  author={Dunne, Gerald V},
  booktitle={From Fields to Strings: Circumnavigating Theoretical Physics: Ian Kogan Memorial Collection (In 3 Volumes)},
  pages={445--522},
  year={2005},
  publisher={World Scientific}
}

@article{Hu98,
    author = "Hu, Wayne",
    title = "{Structure formation with generalized dark matter}",
    eprint = "astro-ph/9801234",
    archivePrefix = "arXiv",
    reportNumber = "IASSNS-AST-98-5",
    doi = "10.1086/306274",
    journal = "Astrophys. J.",
    volume = "506",
    pages = "485--494",
    year = "1998"
}

@article{KST16,
    author = "Kopp, Michael and Skordis, Constantinos and Thomas, Dan B.",
    title = "{Extensive investigation of the generalized dark matter model}",
    eprint = "1605.00649",
    archivePrefix = "arXiv",
    primaryClass = "astro-ph.CO",
    doi = "10.1103/PhysRevD.94.043512",
    journal = "Phys. Rev. D",
    volume = "94",
    number = "4",
    pages = "043512",
    year = "2016"
}

@misc{HN24,
      title={Cosmological perturbations of a relativistic MOND theory}, 
      author={Jai-chan Hwang and Hyerim Noh},
      year={2024},
      eprint={2410.10205},
      archivePrefix={arXiv},
      primaryClass={gr-qc},
      url={https://arxiv.org/abs/2410.10205}, 
}

@article{EulerHeisenberg,
  title={Consequences of dirac theory of the positron},
  author={Heisenberg, W and Euler, H},
    journal = "Z. Phys.",
    volume = "98",
    number = "11-12",
    pages = "714",
    year = "1936",
  journal={arXiv preprint physics/0605038}
}

@article{Finster_2024,
   title={Theoretically Motivated Dark Electromagnetism as the Origin of Relativistic Modified Newtonian Dynamics},
   volume={10},
   ISSN={2218-1997},
   url={http://dx.doi.org/10.3390/universe10030123},
   DOI={10.3390/universe10030123},
   number={3},
   journal={Universe},
   publisher={MDPI AG},
   author={Finster, Felix and Isidro, José M. and Paganini, Claudio F. and Singh, Tejinder P.},
   year={2024},
   month=mar, pages={123} }

@article{PhysRevLett.63.2333,
  title = {Larger Higgs-boson-exchange terms in the neutron electric dipole moment},
  author = {Weinberg, Steven},
  journal = {Phys. Rev. Lett.},
  volume = {63},
  issue = {21},
  pages = {2333--2336},
  numpages = {0},
  year = {1989},
  month = {Nov},
  publisher = {American Physical Society},
  doi = {10.1103/PhysRevLett.63.2333},
  url = {https://link.aps.org/doi/10.1103/PhysRevLett.63.2333}
}

\end{document}